\documentclass[prd,twocolumn,aps,preprintnumbers,letterpaper]{revtex4}
\usepackage{amsmath}
\usepackage{tipa}
\usepackage{amssymb}
\usepackage{graphicx}
\usepackage{bm}
\usepackage{enumerate}
\usepackage{color}
\newcommand{\ddd}{\displaystyle}
\newcommand{\dd}{\mathrm{d}}
\newcommand{\be}{\begin{equation}}
\newcommand{\ee}{\end{equation}}
\newcommand{\bea}{\begin{eqnarray}}
\newcommand{\eea}{\end{eqnarray}}
\newcommand{\ba}{\begin{eqnarray}}
\newcommand{\ea}{\end{eqnarray}}

\newcommand{\GeV}{\mathrm{GeV}}
\newcommand{\fm}{\mathrm{fm}}

\begin{document}

\title{Self-interacting QCD strings and string balls}
\author{Tigran Kalaydzhyan}
\affiliation{Department of Physics and Astronomy, Stony Brook University,\\ Stony Brook, New York 11794-3800, USA}

\author{Edward Shuryak}
\affiliation{Department of Physics and Astronomy, Stony Brook University,\\ Stony Brook, New York 11794-3800, USA}

\date{\today}

\begin{abstract}
Strings at $T\approx T_c$ are known to be subject to the so-called Hagedorn phenomenon, in which a string's entropy
(times $T$)
   and energy cancel each other and result in the evolution of the string into highly excited states, or ``string
balls". Intrinsic attractive interaction of strings -- gravitational for fundamental strings or in the context of holographic models
of the AdS/QCD type, or $\sigma$ exchanges for QCD strings -- can significantly modify properties of the string balls.
If heavy enough, those start approaching properties of the black holes.
We generate self-interacting string balls numerically,
in a thermal string lattice model. We found that in a certain range of the interaction coupling constants
they morph into a new phase, the ``entropy-rich" string balls.
These objects can appear in the so-called mixed phase
of hadronic matter, produced in heavy ion collisions, as well as possibly in the high multiplicity proton-proton or proton-nucleus collisions.
Among discussed applications are jet quenching in the mixed phase and also the study of angular deformations
of the string balls.
 \end{abstract}


\maketitle
\section{Introduction}
\subsection{Overview}
     The
  first hints for the existence of the stringy objects in strong interactions were found in the 1960s, way before  QCD, quarkonia, and their linear potentials between color charges: they came
   from the Regge  phenomenology and the Veneziano scattering amplitude.
   Theoretical attempts to derive membranes (string worldhistories) starting from the perturbative Feynman diagrams were inconclusive, even though such diagrams become planar, or of ``fishnet" kind, in the 't Hooft's large $N_c$ limit.
Only with the advent of AdS/CFT correspondence did
the gauge-string duality become exact for some gauge theories, but alas not (yet) for QCD.

    An important role of the QCD strings at finite temperatures
    stems from the fundamental fact that strings, unlike particles, have an exponentially growing density of states \cite{FV,HV}.
  A decade later it  was noticed by Polyakov \cite{Polyakov:1979gp} and Susskind  \cite{Susskind:1979up}
 that the so-called Hagedorn phenomenon with strings is at the heart of the (strong first-order) deconfinement phase transition
 in the (pure) gauge theories.
 As the string entropy (times $T$) and energy are approximately canceling each other,
 one can get small free energy and as a result
    highly  excited strings.  Those may form
the    {\em string balls}, which are the subject of this paper.

    Historically, studies of the {\em self-interacting} string balls started in the framework
of fundamental string theory: the theoretical questions discussed were related to the understanding of the transition from the string balls to black holes.
We briefly recall the main points of that in Sec.~\ref{sec_to_bh} below.

 Highly excited  strings  populate the
    so-called ``mixed phase" of gluodynamics at $T=T_c$ and provide an energy/entropy density interpolating between the two values
    $\epsilon_{min},\epsilon_{max}$ corresponding to  hadronic and QGP phases. QCD with dynamical quarks has a
    crossover transition, in which the mixed phase can also be defined as a narrow strip of temperatures in which similar evolution happens.
  The so-called ``resonance gas" models used to describe it are consistent with the Hagedorn picture, since the hadronic density of states
  is in good correspondence with those of QCD strings. For lattice studies of this see, e.g., Ref.~\cite{Caselle:2011fy}.

    If the energy of heavy ion collisions is tuned appropriately, such conditions can occur as the $initial$ state of matter produced:
    it is related to the so-called ``softest point" of the equation of state, see, e.g., the discussion of it 20 years ago \cite{Hung:1994eq}.
    However, in
  contemporary ``mainstream" experiments with heavy ion collisions, at Brookhaven Relativistic Heavy Ion Collider and CERN LHC collider,
  the mixed phase appears  as an $intermediate$ condition, between the initial QGP stage and the final hadronic one.
  Still,  between 1/3 and 1/4 of the total evolution (proper) time of the fireball is spent crossing through  the mixed phase; thus, its properties are
  important to understand.
  It is particularly important for certain observables such as  jet quenching
    \cite{Liao:2008dk}.   The string balls have been recently discussed in a completely different context, as initial states for the high
multiplicity $pp$ collisions: we briefly introduce those ideas in Sec.~\ref{sec_pomeron}.

The objective of this paper is to study the role of self-interaction of the string balls, specifically for QCD strings at  $T\approx T_c$,
which is numerically close to the Hagedorn temperature $T_H$. We will formulate a new lattice model for those and
simulate numerically ensembles of string balls of various sizes, using space-dependent temperature $T(x)$.
The main physics issue studied is the dependence of the string balls on the self-interaction coupling.
As we found in Sec.~\ref{sec_with_interaction}, there are two radical changes at certain values of this coupling:
first, a new regime appears, which we call the entropy rich regime; second, the ball undergoes a collapse.
Applications of those results include a section on jet quenching in the mixed phase, Sec.~\ref{sec_quenching}, in which we point out that
current estimates for the jet quenching parameter can be an order of magnitude enhanced, and
Sec.~\ref{sec_angular}, in which we study shape fluctuations of the string balls.
The paper is summarized in Sec.~\ref{sec_summary}; further directions of research are discussed in Sec.~\ref{sec_outlook}.

\subsection{From strings to black holes} \label{sec_to_bh}

   Historically, the subject of string self-interaction was first discussed in the context of fundamental strings in
   critical dimensions (26 for bosonic strings and 10 for superstrings). The string coupling $g_s$ in this case is a function of the vacuum expectation value of the dilaton field, $\phi$: $g_s = e^\phi$ for closed strings and $g_s = e^{\phi/2}$ for open strings. The power of $g_s$ in the string amplitude is then given by the Euler characteristic $\chi$ of the string worldsheet.
   As it is well known, the
   massless modes of closed strings include gravitons; therefore, it is a candidate for the theory of quantum gravity.
     The subject relevant for this work is the transition between the
states of massive string balls and the ones of black holes.    When any object gets very massive, one expects it to be described classically.
 Sufficiently massive string balls should thus become black holes of the classical gravity.

A string ball can be naively generated by  a ``random walk" process,   of $M/M_s$ steps, where $M_s\sim 1/\sqrt{\alpha'}$
  is the typical mass of a straight string segment. If so, the string entropy  scales as the number of segments
  \be S_{ball}\sim M/M_s
  \ee

  The Schwarzschild radius of a black hole in $d$ spatial dimensions is
  \be \ddd R_{BH}\sim \left( M\right) ^{1 \over (d-2)} \label{R_BH} \ee
   and the Bekenstein entropy
   \be \ddd S_{BH}\sim {Area }\sim M^{d-1\over d-2} \ee
 Thus, the equality $S_{ball}= S_{BH}$ can  only be reached
   at some special critical mass $M_c$.  When this happens, the Hawking temperature of the black hole
   is  exactly the string Hagedorn value $T_H$ and the radius is at the string scale.
   So, at least at such a value of the mass, a near-critical string ball
   can be identified -- at least thermodynamically -- with a black hole.

   However, in order to understand how exactly this state is reached, one should first address the
   following puzzle.  Considering a free string ball (described by the Polyakov's near-critical  random walk), one would estimate its radius to be
   \be {R_{ball,r.w.} \over l_s} \sim \sqrt{ M }
   \ee
for any dimension $d$.  This  answer does not fit the Schwarzschild radius $ R_{BH}$ given above (\ref{R_BH}).

The important element  missing is the self-interaction of the string ball: perhaps, Susskind was the first who pointed it out.
 A more quantitative study started by
 Horowitz and Polchinski \cite{Horowitz:1997jc} had used the mean field approach,  and then Damour and Veneziano \cite{Damour:1999aw}
completed the argument by using the correction to the ball's mass due to the self-interaction. Their reasoning can be summarized by
 the following schematic expression for the entropy of a self-interacting  string ball of radius $R$ and mass $M$,
\be
S(M,R)\sim M\left(1-{1\over R^2}\right) \left(1-{R^2 \over M^2}\right)\left(1+{ g^2 M \over R^{d-2}}\right)
\ee
where all numerical constants are for brevity suppressed and all dimensional quantities are in string units
given by its tension.
The coupling
$g$ in the last bracket is the string self-coupling constant to be much discussed below.
 For a very weak coupling, the last term in the last bracket can
be ignored and the entropy maximum will be given by the first two terms; this brings us back to the random walk string ball. However, even
for a very small $g$, the importance
of the last term depends not on $g$ but on $g^2M$. So, very massive balls can be influenced by a very weak
 gravity (what, indeed, happens with planets and stars).
If the last term is large compared to 1, the self-interacting string balls become much smaller in size and eventually fit the Schwarzschild radius.

  \begin{figure}[t]
  \begin{center}
  \includegraphics[width=6cm]{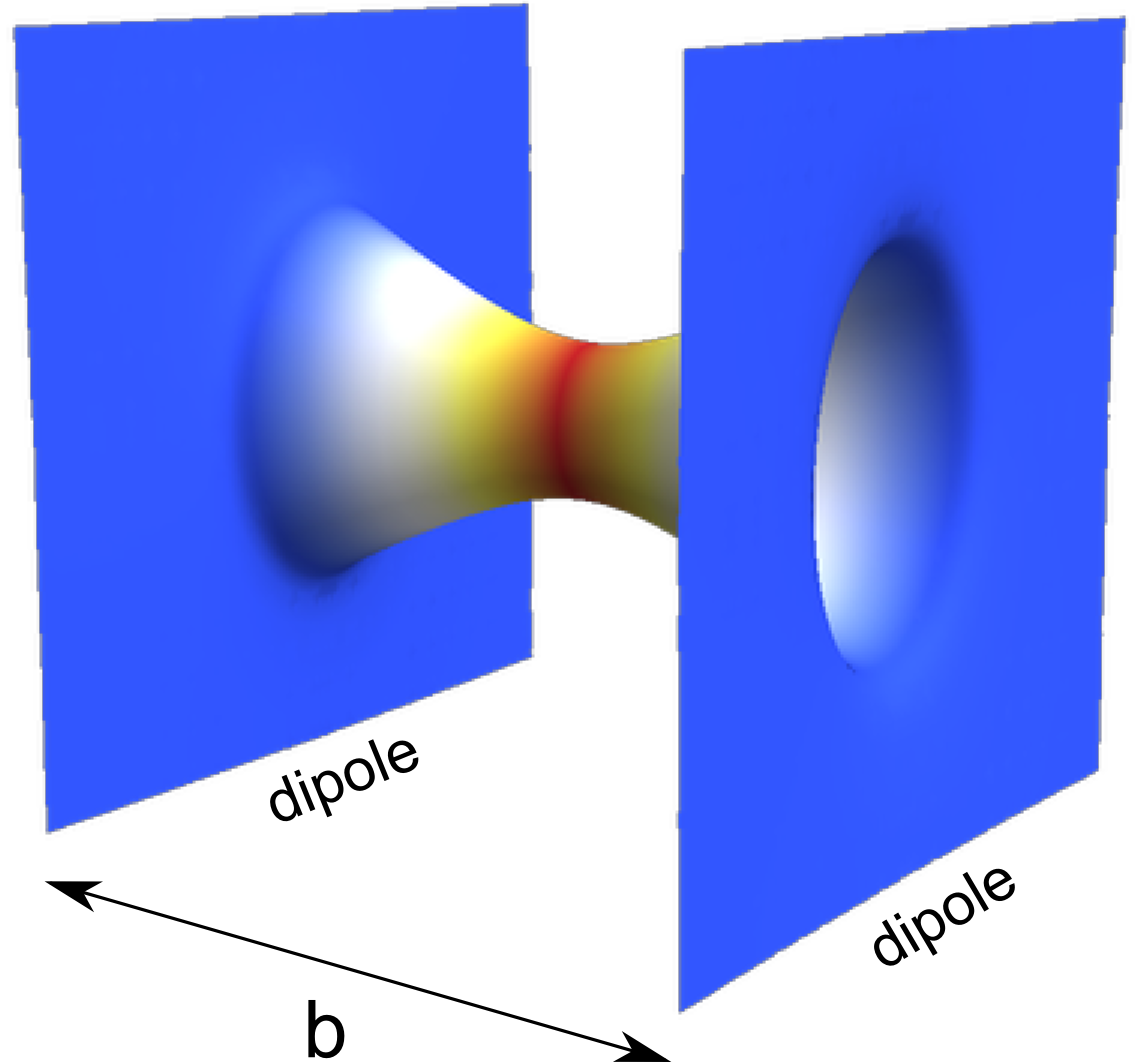}
  \caption{Dipole-dipole scattering due to the tube string configuration. The impact parameter  ${\bf b}$
  is the dipole transverse separation.}
  \label{fig_DD}
  \end{center}
\end{figure}

\subsection{String balls emerging in high-energy $pp$ scattering} \label{sec_pomeron}

The Pomeron description of the high-energy hadronic scattering includes production of (two) QCD
strings stretched between the receding color dipoles. Zahed and collaborators \cite{Basar:2012jb,Stoffers:2012zw}
proposed a semiclassical derivation of the tunneling (Euclidean) stage of the process, based on the
so-called ``tube" string configuration shown schematically in Fig.~\ref{fig_DD}. Depending on how it is cut, it can be viewed as
either a production of two open strings or a closed string exchange between the two color dipoles.

It was recently pointed out in Ref.~\cite{Shuryak:2013sra} that the Hagedorn phenomenon and the resulting string balls
naturally appear in this setting.  The first step is to
recognize that the tube geometry of the surface naturally leads to a periodic coordinate and thermal description:
 the  circumference of the
tube is identified with the Matsubara time $\tau=1/T$ , inverse to the
 effective string temperature.   Furthermore, the tube profile is not uniform along the tube, and, therefore, the temperature $T$ depends on the
  longitudinal coordinate on a tube that we call $\sigma_W \in (0,1) $,
\be
T(\sigma_W)={\chi \over 2 \pi { b}} {1 \over {\rm cosh}(\chi(\sigma_W-1/2))}\,, \label{T_versus_sigma}
\ee
with its highest value at the center or $T(\sigma_W=1/2)\equiv T=\chi/2\pi{ b}$. The parameter $\chi=\ln(s/s_0)$
corresponds to rapidity of the colliding dipoles, which in Euclidean formulation becomes the usual angle
between them. In the string diffusion equation, it
also has the meaning of the diffusion time; so with growing $s$ this time increases, the string tube gets longer,
and the cross section grows.

 The second step is to note that at certain values of the
 impact parameter $b$ this temperature
corresponds to the Hagedorn value; the effective tension of the string decreases, and
its high excitations become possible. As a result, as one can expect (and, indeed, sees it directly in the
observed elastic scattering profile), the scattering amplitude for such $b$ exceeds the value
interpolated by a Pomeron string expression from large $b$. One also finds an abrupt change to the
$b$-independent profile at smaller $b$, interpreted in Ref.~\cite{Shuryak:2013sra} as the end of the mixed phase
and transition to the deconfined (or black hole) phase.
Such interpretation suggests prompt
production of the string balls in a mixed phase,
 between ``cold string" Pomerons at large $b$ and a perturbative  domain
at small $b$. High entropy of these balls may lead to very high multiplicity
events observed.

  It is important to emphasize that the temperature here is just the effective
  description of quantum string excitations in the Euclidean partition function under the barrier.
 String balls we discuss in this section therefore appear instantaneously, at $t=0$, as they
emerge in the  (Minkowski or real time) part of the system's path, from the Euclidean  tunneling path.
In distinction to heavy ion collisions,
 no time is needed for this ``thermalization".

While in this paper we will discuss QCD strings in the $d=3$ spatial dimension, corresponding to the mixed phase of QCD, we also would like to
keep in mind that at time zero the ``prompt" near-critical string balls in $pp$ collisions
should not be very different.
In particular, the lack of one coordinate -- the system has near-zero size along the beam direction -- is compensated by the presence of an
extra holographic direction $z$, so it is three-dimensional. (Of course,   $z$ is curved and has some end, so it is not exactly the same as another spatial dimension. Also the self-interaction is a bit
different.  But we do not think that those effects modify the main physics too much.)
What is very different is the fate of such string balls: after $t=0$ they are violently stretched
along the beam directions, as the ends of the strings are still attached to the beam particle fragments,
moving with a large rapidity.

\section{Self-interacting strings}

\subsection{Self-interaction and nuclear physics} \label{sec_interaction1}

For the purpose of this first  qualitative  study, we focus only on the lightest scalar state,
known in hadronic phenomenology as the $\sigma$ meson, or $f_0(500)$ in the PDG13 listings.
Its mass $m_\sigma=  0.4-0.55\, \GeV$  is comparable to its width   $\Gamma_\sigma=  0.4-0.7\, \GeV$; that
is one of several reasons why this mesonic $J^{PC}=0^{++}$ resonance has a difficult history, appearing and disappearing in the Particle Data Group tables.
 The interpretation of its parameters and its dynamical origin has been varying as well, as arguments on this subject have not yet converged. Avoiding the debate we fix the mass in vacuum to be
 $m_\sigma=  0.6\, \GeV$  and assume zero width.
 Below we will also use variable mass $T$-dependent mass, and discuss its coupling to QCD strings.
We will use ``dilatonic" notations,  as if the interaction is made with $T^{\mu\mu}$.

 For one particle species -- the nucleon $N$ -- its coupling is reasonably well known, as it
 is the main component of the attractive central part of nuclear potential. It takes the Yukawa form
  \be V_{NN}(r) = { g^2_{\sigma NN} \over 4 \pi} { \exp(-m_\sigma r) \over r } \ee
  and is  mostly responsible for the nuclear binding.

 For non-nuclear physicists it may be worth reminding the reader at this point that in the $NN$ case it is nearly completely cancelled by the repulsive
 vector $\omega$ exchanges, coupled to the nucleon baryon number. We also remind the reader that
 this sigma term can be found in phenomenological potentials such as Paris and Bonn ones, or the so-called Walecka model of nuclear forces. More recent treatment uses a more accurate ``correlated $\pi\pi$" exchange
 to account for it.

 For non-string theorists it may be worth reminding that the fundamental strings and D-branes
 have also certain charges and repulsive vector forces, canceling attractive ones and
 making them  ``BPS-protected". Our QCD string is not like that; it is
 just a bosonic string without charges, and there are no traces of supersymmetry or BPS protection.

  \begin{figure}[t]
  \begin{center}
  \includegraphics[width=6cm]{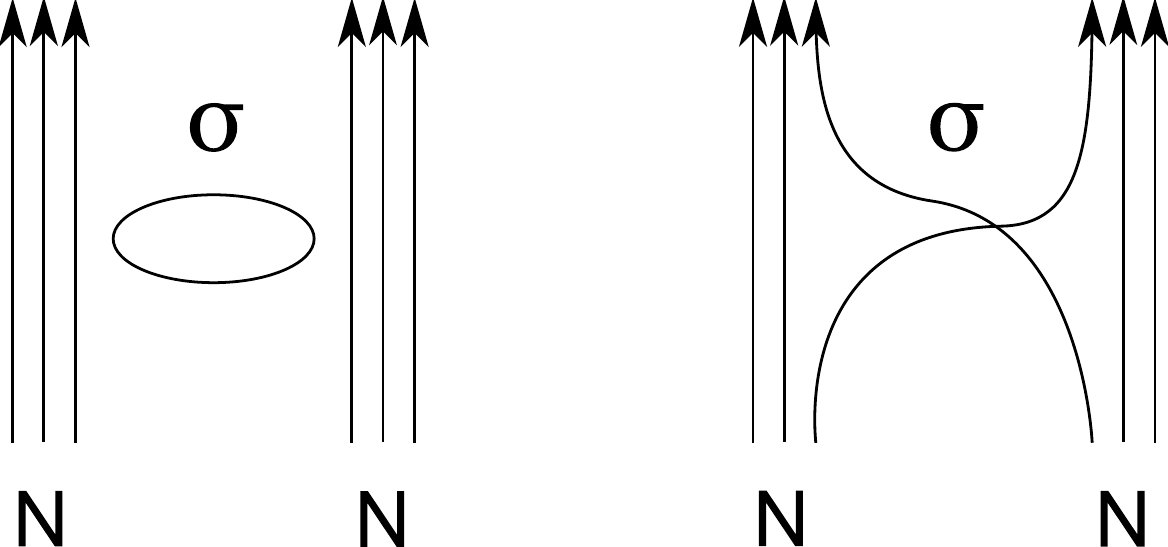}
  \caption{Color diagrams for $\sigma$ exchange}
  \label{fig_diags}
  \end{center}
\end{figure}

  So, one may think of the string balls we study as some
 ``simplified nuclei", for which the Fermi blocking, the repulsive  vector interactions, and forces related to the spin-isospin interaction are all switched out, with only
  $\sigma$-induced attraction left,  binding them together.

   There are two color 't Hooft diagrams (see Fig.~\ref{fig_diags}) contributing to this coupling: the first is suppressed at the limit of a large number of colors, and the second (nonplanar one) is in fact suppressed twice, both in the number of colors and flavors. Obviously, only the first  one can contribute to the self-interaction of strings. Since  we do not know the relative
   contributions of those diagrams, or, alternatively, the fraction of the ``dilaton" in the $\sigma$ meson,
   we take the strength of the $\sigma NN$ coupling as an estimate on its $upper$ limit.
  In terms  of
 the ``scalar Newton's constant" $g_N$ for the self-interaction of QCD strings,  its benchmark value is
 \be g_N^{max} = { g^2_{\sigma NN} \over 4 \pi m_N^2} \approx { 357  \over 4 \pi} {m_\sigma^2 \over m_N^4}\approx 13 \, \GeV^{-2}  \label{sigma_NN} \ee
 In what follows we will
treat $g_N$ as an unknown parameter; in the next subsection,  we will discuss its values indicated by lattice simulations.

The simplest problem in nuclear matter one can consider analytically is the infinite matter with constant (zeroth-order mass) density $\rho_0$. The shift in the energy density
due to the self-interaction is proportional to the space integral of the potential
\be {|\delta \rho | \over \rho_0} = \rho_0 \int \dd^3 x\, V(|x|)={4\pi g_N \rho_0 \over m_\sigma^2 } 
 \ee
Note that the correction diverges for the (gravitylike) massless limit  $g_N=const,m_\sigma \rightarrow 0$,
as the static Universe filled with matter cannot exist. However,
in the last expression, the sigma mass cancels out, and, therefore, the result does not depend on its (rather uncertain)
value. (This happens because  its nuclear parametrization in the form (\ref{sigma_NN}) was  done with
the idea of keeping properties of the nuclear matter independent on it as well).

The previous expression naturally leads to a concept of the critical mass density, at which the negative self-interaction energy
cancels the original zeroth-order mass,  $\delta \rho + \rho_0=0$,
\be \rho_c=\left( {m_N^4 \over 357.}\approx 0.28 {\GeV \over \fm^3}\right)  {g_N^{max} \over g_N}\ee
For the maximal coupling $g_N=g_N^{max}$, it is about twice the mass density for the symmetric nuclear matter $\rho_{n.m.}\approx 0.149 \, \GeV/\fm^3$.
However, the energy density of the mixed phase of interest is in fact up to an order of magnitude higher;
this suggests that the coupling $g_N$ should in fact be substantially smaller than $g_N^{max} $.

At the density $\rho>\rho_c$, it becomes energetically more favorable to produce new string segments.
The process of production stabilizes at the upper high-energy density cutoff of our model. This is a
scenario preceding a
gravitational collapse. On the other hand, we consider our strings
 to be in a contact with a heat bath of a certain temperature; therefore, their stability depends
on a (much stronger)  condition, based on the $free$ energy rather than the energy itself.

 Of course, these arguments only apply to very large systems, much larger than the correlation length
 $m_\sigma^{-1}$, and below we will study finite size string balls.
 We solved some spherically symmetric examples as well, but the results are not particularly
 instructive to be discussed. Our main objective is to study string configurations of an arbitrary shape, which is a suitable task for the numerical simulations.

 Can a collapsing ball be stabilized? One  natural  cutoff  for the
  strings density follows from the self-avoiding rule
  in our lattice model.
One may also wonder, since $\sigma$ is a meson, if its effective Lagrangian includes repulsive nonlinear terms $O(\sigma^4)$, on top
  of its kinetic and
mass terms leading to the Yukawa expression used. We have not studied this option, partly because
in the AdS/QCD setting   --  which we describe shortly -- the nonlinear actions
are well defined and  known, yet the gravitational collapse is unavoidable.

The intriguing feature of a collapsing ball is a continuous production of entropy, resulting from the fact that
very dense string balls have a huge number of (classical) configurations. So, we have our version of the information paradox.
Like evaporating Hawking black holes finally disappearing, the string balls at the mixed phase all eventually decay into
hadrons, as the heavy ion collision reaches its hadronic phase. The
 string entropy stored in these balls should also be eventually released into the final clusters.

\subsection{Self-interaction on the lattice} \label{sec_interaction2}

   While observation of QCD flux tubes was one of the major achievements of the lattice gauge theory,
   unfortunately their interaction has not yet been systematically addressed. This section is a summary of what we were
   able to find in the literature.

   Among the first papers in which the issue has been studied on the lattice was a paper by the Vienna group
\cite{Zach:1997kf} in 1997. Using for technical reasons $U(1)$ lattice gauge theory in 2+1 and 3+1 dimensions, they
observed that in the latter case strong attractive interaction between two parallel single-flux tubes. The effect is strongly enhanced near
the phase transition. In Sec.~6 of their paper, they in particularly study an arrangement of two tubes of length $d=22a$
induced by charges separated by $d=4a$ and show that in the middle two tubes basically merge into one -- see their Fig.~12.
An interesting observation is that the longitudinal electric field retains the same magnitude as in a single tube and changes little
in its total energy: the whole effect comes from the ``coil" provided by a magnetic current around the tubes, which
tends to become one joint coil around both tubes.

  Qualitative influence of the flux tubes on vacuum structure was demonstrated already in early 1990s. In particular,
topological activity -- instantons -- was shown to be suppressed around the string.  From the perspective of this work,
one is naturally interested in ``mesonic clouds" around the flux tubes, and more specifically of the scalar $\sigma$.
To get some quantitative result about it, one obviously needs to perform lattice simulation with dynamical quarks,
which would be light enough and possess good chiral symmetry -- otherwise  $\sigma$ itself is known to get distorted.
Only quite recently the progress in computing performance and development of chiral (overlap) fermions made it possible to do such simulations.

\begin{figure}[t]
  \begin{center}
  \includegraphics[width=6cm]{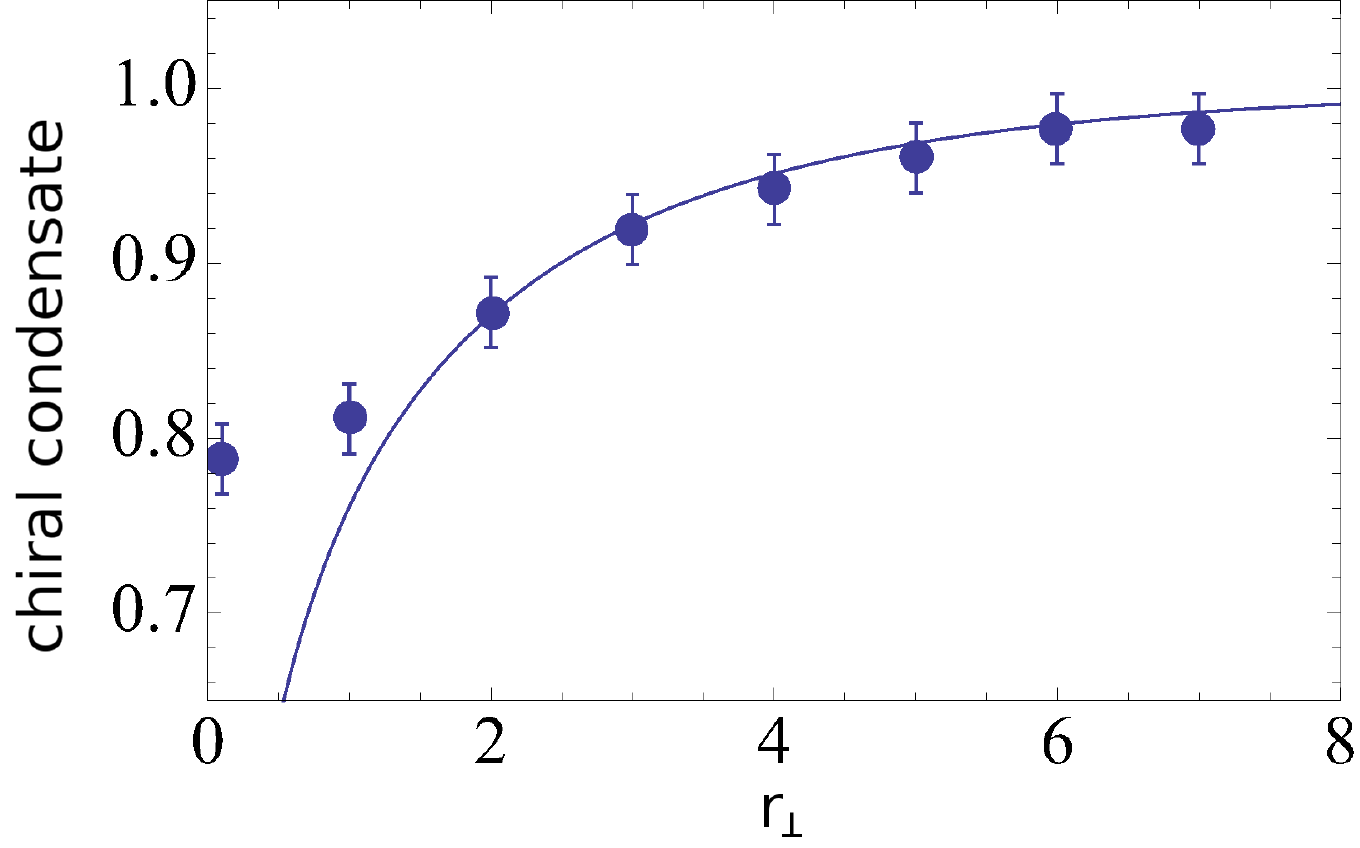}
   \caption{The quantity  $\langle\bar q q(r_\perp) W\rangle/\langle\bar q q\rangle\langle W\rangle$ vs the coordinate $r_\perp$ perpendicular to the flux tube and the Wilson line $W$,
   in units of the lattice spacing. The curve is the fit ($\ref{eqn_cloud})$ with $C=0.21$.
   }
  \label{fig_qq}
  \end{center}
\end{figure}

  One recent study  \cite{Iritani:2013rla}
   at zero
temperature was done using $16^3 \times 48$ lattice, good chiral fermions,
and rather small lattice spacing $a= (1.76\, \GeV)^{-1}=0.11 \, \fm$.
 The authors have  measured the correlation between the flux tube and certain local
observables. Among those there are  two  scalar ones, the squared gluon field strength [filtered with the lowest Dirac eigenvalues;
see the definition in Eq.~(2.4) of that paper] and the chiral condensate correlated with the Wilson line $\langle\bar q q\rangle_W$ creating the flux.  The data points corresponding to the
latter observable are reproduced  in Fig.~\ref{fig_qq}. (Since the shape of the other scalar operator is very similar, we do not show it.)

     A straight infinitely thin
     string being the source of the scalar field with Yukawa potential generates a
  sigma field around it of the magnitude
  \be
   \sigma(r_\perp)= g_N^{1/2} \sigma_T\, 2 K_0(m_\sigma   r_\perp)\,,
  \ee
  where the source is the energy of the string tension and $K_0$ is the modified Bessel function. We have compared the data of Ref.~\cite{Iritani:2013rla}  with the expected shape of the sigma cloud, with $ m_\sigma=600\, \mathrm{MeV}$,
 \be  {\langle\bar q q(r_\perp) W\rangle \over \langle\bar q q\rangle\langle W\rangle}= 1-C\times  K_0(m_\sigma  r_\perp)\,, \label{eqn_cloud} \ee
and found a very good agreement.  So, a sigma cloud around the string is, indeed, seen on the lattice!

(Only the first two points deviate from the curve, indicating that the string is not infinitely thin
but rather
has a width of the order of one lattice spacing.
Note that this small value does not contradict other data on the string size we mentioned above,
because here we discuss the ``bare" string, stripped from its mesonic cloud, not the total effective object.)

A rough
estimate of the strength of string coupling to the $\sigma$ field can now be obtained from the fit (\ref{eqn_cloud})
and the fact that sigma couples to quarks directly, $\sigma \bar{q} q $. The latter means that the quark effective mass is given by the vacuum expectation value of $\sigma$,
\begin{align}
|\langle\sigma\rangle|\approx 0.35\, \GeV\,,
\end{align}
\begin{align}
g_N^{1/2} \sim { |\langle \sigma\rangle| C \over 2\sigma_T} \approx 0.2 \, \GeV^{-1}\,.
\end{align}

This derived interaction strength is, however, not directly relevant to the string balls discussed in this paper, because
it is studied in vacuum at $T=0$, while we need it in the mixed phase at $T=T_c$. This is expected to
radically change the property of the scalar meson and its ``cloud".

 Let us start with an exactly chirally symmetric limit of QCD with two massless quark flavors. In this case we know
 that the chiral phase transition is of the second order and that $\sigma$ should become a chiral partner of the massless $\pi$, and, therefore,
 \be m_\sigma(T \rightarrow T_c) \rightarrow 0\,. \ee
This would have a rather profound effect on the string ball: if attraction gets infinite range, large enough string balls
would collapsed even for modest coupling.

What happens in the real world with massive quarks and pions has been the subject of  many debates.  Some of those
are collected in the proceedings of the  Kyoto workshop \cite{Ishida:2001pb}. The bottom line is that, while the pion mass
does not have strong temperature dependence,  the $\sigma$  is expected to change
drastically, with $m_\sigma$ merging with $m_\pi$ at and even above $T_c$ (provided they still exist as resonances).
The width $\Gamma_\sigma$ is expected to get reduced even more, or even disappear altogether
if $m_\sigma<2m_\pi$.

In summary,  general expectations and some lattice data suggest the attractive self-interaction of QCD strings to be strongly enhanced, as one moves from $T=0$ to the near-critical region, $T=T_c$.
Unfortunately, lattice studies of it had only been done either in models very far from the real-world QCD \cite{Zach:1997kf}
or in a more realistic setting \cite{Iritani:2013rla}, but only at $T=0$. We call upon lattice practitioners to fill this gap; as far as we know, there are no technical reasons not to do so.

\subsection{Self-interaction of holographic strings}
   Although we do not really discuss
 holographic AdS/QCD string balls in this work, let us still comment on those.
 Most of the works on holography are done in the limit $N_c\rightarrow \infty$, to put those into the classical gravity domain,
while the number of quark flavors $N_f$ is considered to be finite. This approximation is similar to the ``quenched" one in lattice gauge theories,
and it ignores the backreaction of quarks on the glue. An analog of simulations with the dynamical quarks in the holographic world
is known as the Veneziano limit   $N_c, N_f\rightarrow \infty$, $N_f/N_c=const$, sometimes called V-QCD.

 In all such approaches there are
 massless fields
in the bulk Lagrangian,  such as the $dilaton$
and the $graviton$. They are interacting with the components of the stress tensor,  $T^{\mu\mu}$ and $T^{\mu\nu}$, respectively,
in a standard manner.
The existence of the confining wall in the holographic direction leads to the quantization  of the
motion in this direction, effectively making propagation in other directions massive. For a specific choice of the wall -- e.g., the so-called ``soft wall" \cite{Karch:2006pv} -- one can easily calculate the mass spectrum of hadrons: typically one gets linear
 Regge trajectories. In this sense, a massless bulk dilaton and graviton correspond
to a whole trajectory  of massive hadrons in the gauge theory.

  As a recent example of a holographic AdS/QCD model, working in the  Veneziano limit, one can take Ref.~\cite{Arean:2013tja},
  in which holographic dual gravity solution is developed. What is more relevant for us, is that in this work
the masses of scalar hadrons are calculated as a function of $N_f$. In Fig.~7 of that work, one finds such behavior
 for the four lowest scalar (flavor single) states:  the lowest is the $\sigma$ meson, the next is the ``scalar glueball",
 and one excited state of each species.  The mass ratio of the first pair is $m_{glueball}/m_\sigma\approx 2.5$,
close to the ratio in the real world. An extension of this calculation
in V-QCD for $finite$ temperatures is not done yet (but should be done). The authors of Ref.~\cite{Arean:2013tja} also focused on
the transition to the conformal regime at critical $N_f/N_c\approx 4$. As one can see from their results, near this transition the sigma mass $m_\sigma/\Lambda_{QCD}$ rapidly drops, too. As all holographic models, that one contains also the analog of the flux tube -- the fundamental string in the bulk.
It would be interesting to calculate in this model the strength and range of the string self-interaction.

Now we comment on the dimensionality of the string ball produced by the holographic Pomeron.
Specific dynamics of a high-energy collision leads to the near-vanishing values of time and longitudinal (beam) coordinates $x^0,x^1\approx 0$,
so only the ``transverse" coordinates $x^2,x^3,x^5=z$ are left for string fluctuations. As a result, the effective space is also three dimensional, as for
the usual QCD strings in  space. The difference comes from the metric curved in the $z$ direction.

Even though the subsequent evolution of the string ball at $t>0$ has not yet been studied, the gravitational language
of the holography allows us to introduce the notion  of the ``trapped surface". It can be calculated instantaneously, at $t=0$.
It may or may not exist for a given matter distribution; for example, in holographic collisions with a nonzero impact
parameter, there is a critical value $b_c$ above which the trapped surface disappears  \cite{Lin:2009pn,Gubser:2009sx}.

 \begin{figure}[t]
  \begin{center}
  \includegraphics[width=6cm]{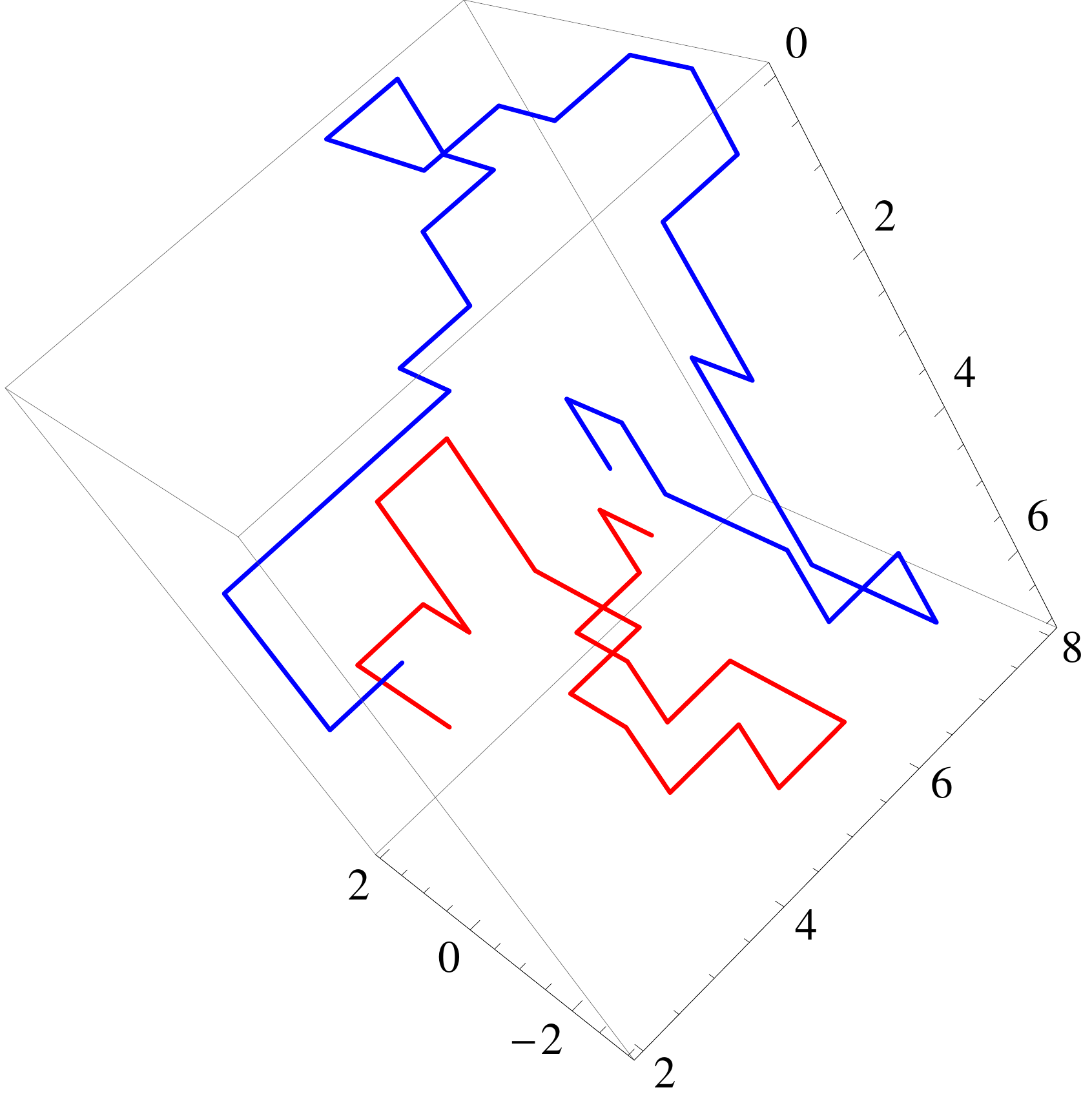}
  \caption{Example of a two-string configuration (a sparse string ball); two strings are plotted as blue and red.}
  \label{fig_shape}
  \end{center}
\end{figure}

\section{Thermal strings on the lattice } \label{sec_lattice}
After our extensive introduction, we introduce the numerical model we use to study the
string balls with self-interaction. While we discuss the details of the setting below in this
section, let us emphasize on the onset its main physics prerequisites, namely, that
the ball surface should be approximately near the Hagedorn temperature, making the
string fluctuate widely outward.

  Following a bit Wilson's strong coupling expansion,
we place the strings on links of a $(d=3)$-dimensional lattice.  Strings are assumed to be in contact with a
heat bath, and a partition function includes all possible string configurations.

Intersections of the strings are not included because
of the repulsive interaction at small distances. Even for the Abelian fields, which
add up simply as vectors,
the action is quadratic in fields (no commutators), and intersections are energetically not favorable.
An exception (in the lattice geometry) is the case of exactly oppositely directed fluxes,
when a part of the string should basically disappear.
We had not included this complication believing that the total entropy and
energy of the string ball will not be affected much.

Instead of using boxes (with or without periodic boundary conditions) as is customary in the lattice gauge
theory and many other statistical  applications, we opted for an infinite space (no box).  Instead the temperature $T$ is {\em space
dependent}. We think it better corresponds to the experimental situation. Furthermore, the
string ball surface is automatically near criticality and thus strongly fluctuating;
this aspect will be important for our application of initial deformations below.

The ``physical units" in gluodynamics, as in lattice tradition, are set by putting
the string tension to its value in the real world:
\be \sigma_T = (0.42\, \GeV)^2 \ee
Numerical lattice simulations
have shown that gluodynamics with $N_c>2$ has a first-order
deconfinement phase transition, with $T_c/\sqrt{\sigma_T}$ very weakly dependent
on $N_c$  (for review, see, e.g., Refs.~\cite{Teper:2009uf, Lucini:2012gg}).  Numerically, the critical temperature of the gluodynamics is $T_c\approx 270\, \mathrm{MeV}$.

It has been further shown that the effective string tension of the free energy $\sigma_{F}(T)$ decreases with $T$; a point
where it vanishes is known as the Hagedorn point. Since this point is above $T_c$, some
attempts have been made \cite{Bringoltz:2005xx} to get closer to it by ``superheating" the
hadronic phase, yet some amount of extrapolation is still needed. The resulting value was found to be
\be {T_H \over T_c}= 1.11    \label{eqn_T_H} \ee

The nature of the lattice model we use is very different from that of the lattice gauge theory (LGT).
First of all, we do not want to study quantum strings and generate two-dimensional surfaces in the Matsubara $R^d S^1$
space, restricting ourselves to the thermodynamics of strings in $d$ spatial dimensions.

The lattice spacing $a$ in LGT is a technical cutoff, which at the end of the calculation is expected to
be extrapolated to zero, reaching the so-called continuum limit. In our case $a$ is a physical parameter
characterizing QCD strings: its value is selected from the requirement that it determines the
correct density of states.
Since we postulate that the string can go to any of $2d-1$ directions from each point
(going backward on itself is prohibited), we have $(2d-1)^{L/a}$ possible strings of length $L$.
Our partition function is given by
\be Z\sim \int \dd L \exp \left[ {L \over a} \ln( 2d-1)- {\sigma_T L \over T} \right]\,,
\ee
and hence the Hagedorn divergence happens at
\be T_H={\sigma_T a \over  \ln( 2d-1)}. \ee
Setting $T_H=0.30\, \GeV$, according to the lattice data mentioned above and the string tension,
we fix the three-dimensional spacing to be
\be a_3= 2.73 \, \GeV^{-1} \approx 0.54\, \fm. \ee
It is, therefore, a much more coarse lattice, compared to the ones usually used in LGT.

If no external charges are involved, the excitations are closed strings.
At low $T$ one may expect to excite only the smallest ones. With the ``no self-crossing" rule we apply,
that would be an elementary plaquette with four links. Its mass,
\be E_{plaquette}=4\sigma_T a\approx 1.9 \, \GeV \,,\ee
is amusingly in the ballpark of the lowest glueball masses of QCD.
(For completeness, the lowest ``meson" is one link or mass 0.5 $\GeV$, and the lowest ``baryon" is three links -- $1.5\, \GeV$ of string energy --  plus that of the ``baryon junction".)

At temperatures below and not  close to $T_H$, one finds extremely dilute $\mathcal{O}(e^{-10})$
gas of glueballs, or straight initial strings we put in. Only close to $T_H$ do multiple string states get excited;
the strings rapidly grow and start occupying a larger and larger fraction of the available space.

Before we show the results of the simulation, let us discuss the opposite ``dense" limit of our model.
We do not allow strings to overlap; the minimal distance between them is one link length, or again about $0.5 \, \fm$.
Is it large enough for the string to be considered well separated? We think so, as it is about three times the string radius
[see discussion below around Eq.~(\ref{eqn_profile})].

 \begin{figure}[t]
  \begin{center}
  \includegraphics[width=7cm]{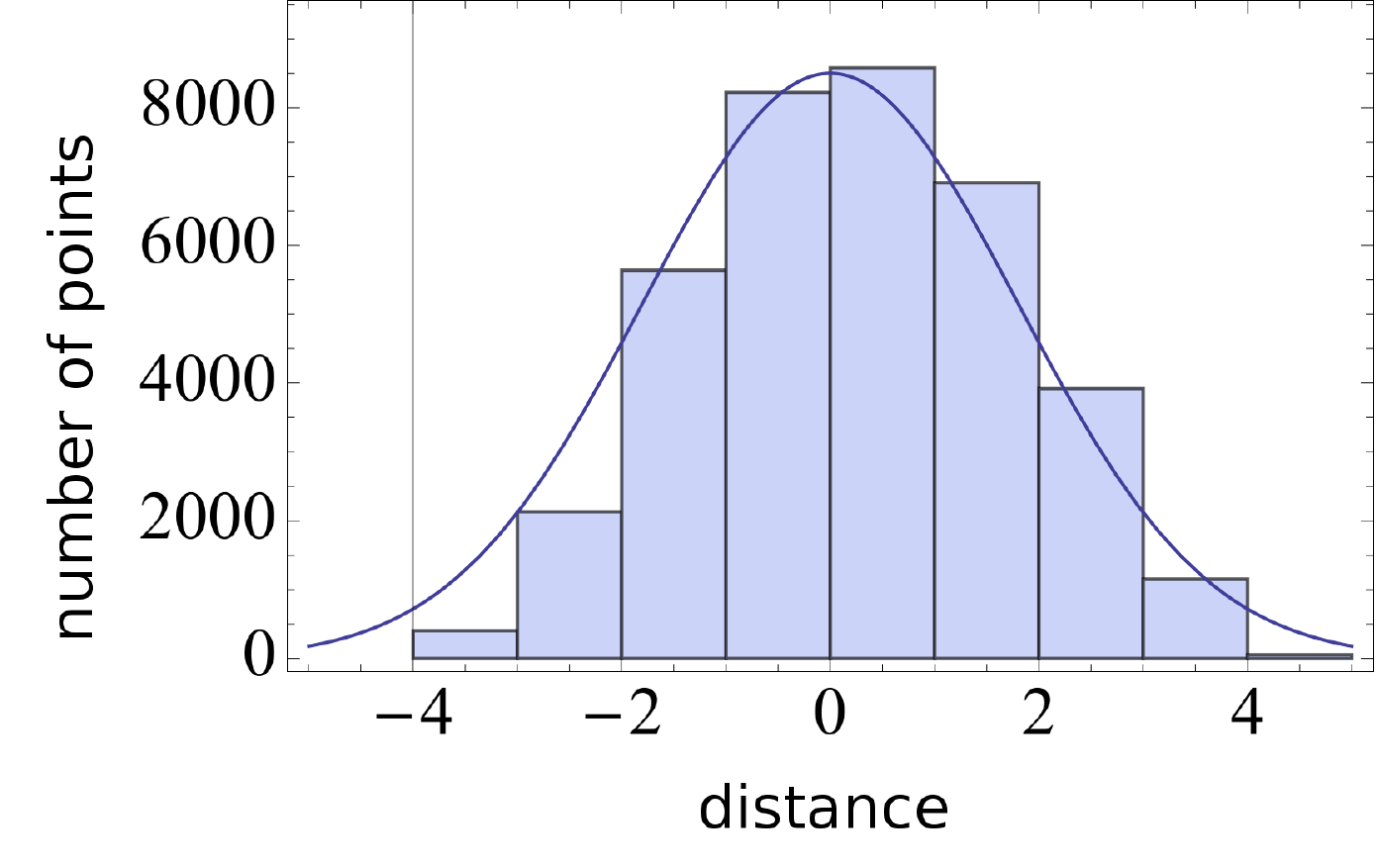}\vspace{3mm}
      \includegraphics[width=7cm]{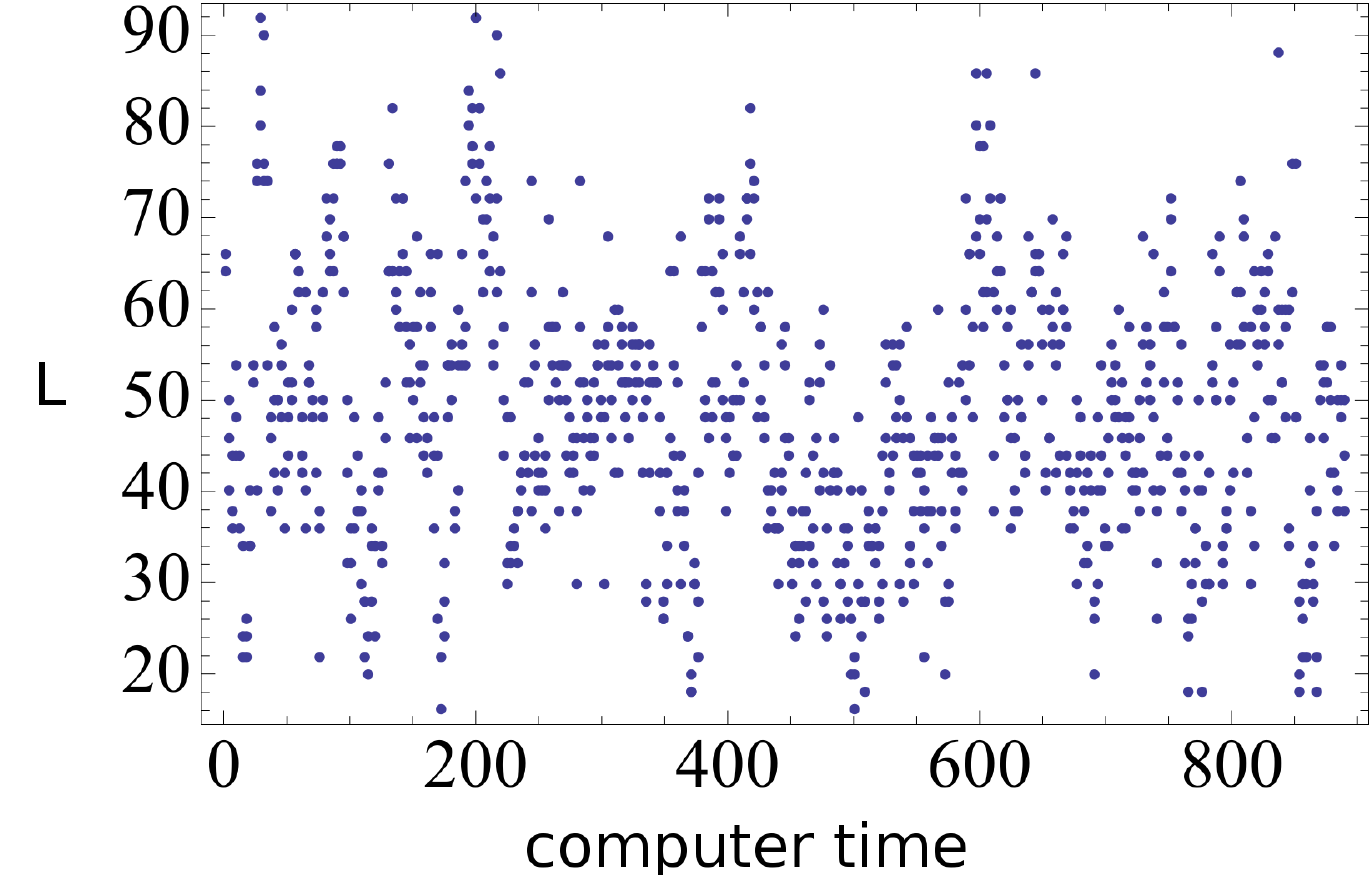}
  \caption{ Upper plot: distribution of all points through all the ensembles along one of the spatial coordinates (in lattice units), compared to
  a Gaussian distribution. Lower plot: dependence of the string length $L$ (lattice units) on the computer time $t$ (in units of 10 full iterations).  Both for the
  zero self-coupling and $T_0=1\, \GeV,\, s_T=2a\approx 1 \, \fm$ simulation.}
  \label{fig_L(t)_g0}
  \end{center}
\end{figure}

The most compact (volume-filling or Hamiltonian) string wrapping visits each site of the lattice. If the string is closed, then the number of occupied links is the same as the number of occupied sites. Since in $d=3$ each site is shared among eight neighboring cubes, there is effectively only one occupied link per unit cube, and this wrapping produces the maximal energy density,
\be {\epsilon_{max} \over T_c^4} = {\sigma_T a \over a^3 T_c^4} \approx 4.4 \ee
(we normalized it to a power of $T_c$, the highest temperature of the hadronic phase).
It is instructive to compare it to the energy density of the gluonic plasma, for which we use the free Stefan-Boltzmann
value
\be   {\epsilon_{gluons} \over T^4}=(N_c^2-1) {\pi^2 \over 15}\approx
5.26 \ee
and conclude that our model's maximal energy density  is comparable to the physical maximal energy density
of the mixed phase we would like to study.

One remaining issue is treatment of color number. In practice we
ignore it, considering thermal excitations of two strings we always initiate the system with.
We also think of those strings as direct and reverse color fluxes from two neutral hadrons, which appear in hadronic collisions;
it basically means that all our strings have only $one$ and the same color.
 Their mutual repulsion
-- or no-crossing rule -- is in this case natural.
All we simulate is
the Hagedorn phenomenon due to the exponentially large number of string states, ignoring prefactors due to the $N_c$.

Some  justification for that comes from the fact that (apart from the properties of the deconfined phase itself)
very little $N_c$ dependence is seen in the lattice gluodynamics data; for a review, see Refs.~\cite{Teper:2009uf,Lucini:2012gg}.
One may, however, still wonder if one should assign specific colors to strings in the model and account for the fact that
two overlapping  flux tubes with  $different$  colors may be partially allowed. In this first study, we simply did not want to
make our model too complex.

\section{Numerical simulations}
\subsection{String ball without a self-interaction }

Our algorithm consists of a sequence of updates for the each string segment, such that the configuration gradually approaches equilibrium.
To thermalize the string and to generate a statistical ensemble, we use the following three types of elementary updates:
\be
\setlength{\unitlength}{.015in}
\begin{picture}(50,13)(0,17)
  \put(10,10){\circle*{3}}
  \put(10,10){\line(0,1){20}}
  \put(10,30){\circle*{3}}
  \put(10,30){\line(1,0){20}}
  \put(30,30){\circle*{3}}
  \put(30,30){\line(0,-1){20}}
  \put(30,10){\circle*{3}}
\end{picture}
\longrightarrow\quad
\begin{picture}(50,13)(0,17)
  \put(10,10){\circle*{3}}
  \put(10,10){\line(1,0){20}}
  \put(30,10){\circle*{3}}
\end{picture}
\ee
\be
\setlength{\unitlength}{.015in}
\begin{picture}(50,13)(0,17)
  \put(10,10){\circle*{3}}
  \put(10,10){\line(0,1){20}}
  \put(10,30){\circle*{3}}
  \put(10,30){\line(1,0){20}}
  \put(30,30){\circle*{3}}
\end{picture}
\longleftrightarrow\quad
\begin{picture}(50,13)(0,17)
  \put(30,30){\circle*{3}}
  \put(30,30){\line(0,-1){20}}
  \put(30,10){\circle*{3}}
  \put(30,10){\line(-1,0){20}}
  \put(10,10){\circle*{3}}
\end{picture}
\ee
\be
\setlength{\unitlength}{.015in}
\begin{picture}(50,13)(0,17)
  \put(10,10){\circle*{3}}
  \put(10,10){\line(1,0){20}}
  \put(30,10){\circle*{3}}
\end{picture}
\longrightarrow\quad
\begin{picture}(50,13)(0,17)
  \put(10,10){\circle*{3}}
  \put(10,10){\line(0,1){20}}
  \put(10,30){\circle*{3}}
  \put(10,30){\line(1,0){20}}
  \put(30,30){\circle*{3}}
  \put(30,30){\line(0,-1){20}}
  \put(30,10){\circle*{3}}
\end{picture}
\vspace{0.4cm}
\ee
There is no 1 to 2 links, because those are ``local  updates",  done with the ends fixed,
where the new ``corners'' and ``staples'' are chosen in a way avoiding self-intersections. A new configuration is then accepted with the probability from the heat bath (Metropolis) algorithm,
\begin{align}
 P_A = \min \left[1, \exp\left(\frac{E_{\mathrm{old}}-E_{\mathrm{new}}}{T}\right) \right]\,,\label{metropolis}
\end{align}
where ($E_{\mathrm{old}}$) $E_{\mathrm{new}}$ is the total energy of the (old) new configuration and $T$ is the temperature in the region of space, where the update is performed.
We introduce a space-dependent temperature with a Gaussian  profile:
\be T(r)= T_{0}\, \exp\left(-{ r^2 \over 2 s_T^2}\right) \ee
As the self-interaction is absent ($g_N=0$), the physics is simple: in the ``cold" regions of space $T(x)<T_H$, the string's entropy times temperature
is less than its energy, and the string segments are only present if they should cross the region in order to connect fixed string ends. In the ``hot" region, where $T(x)>T_H$, the string gets strongly excited.

 \begin{figure}[b]
  \begin{center}
  \includegraphics[width=7cm]{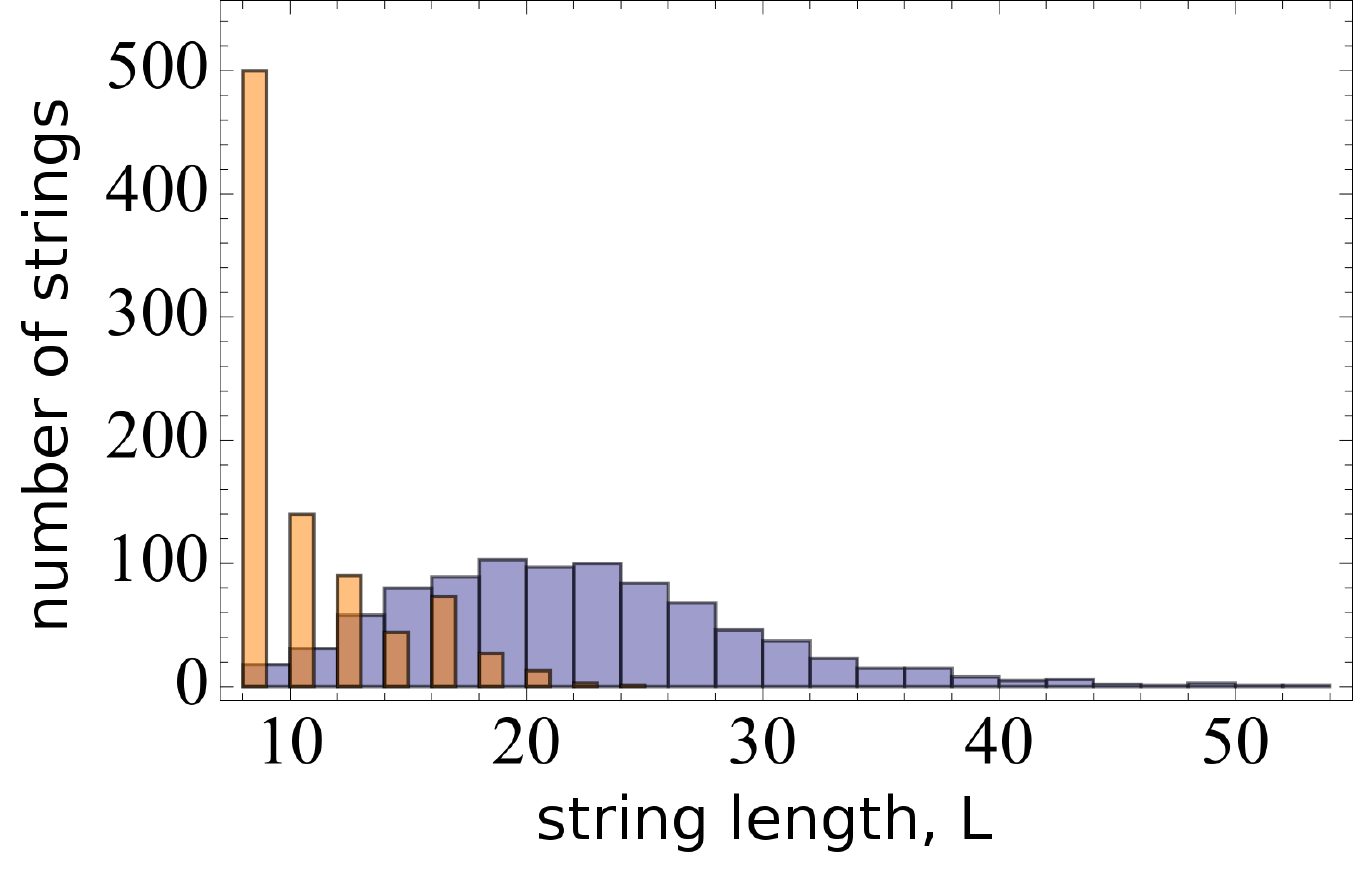}\vspace{3mm}
    \includegraphics[width=5cm]{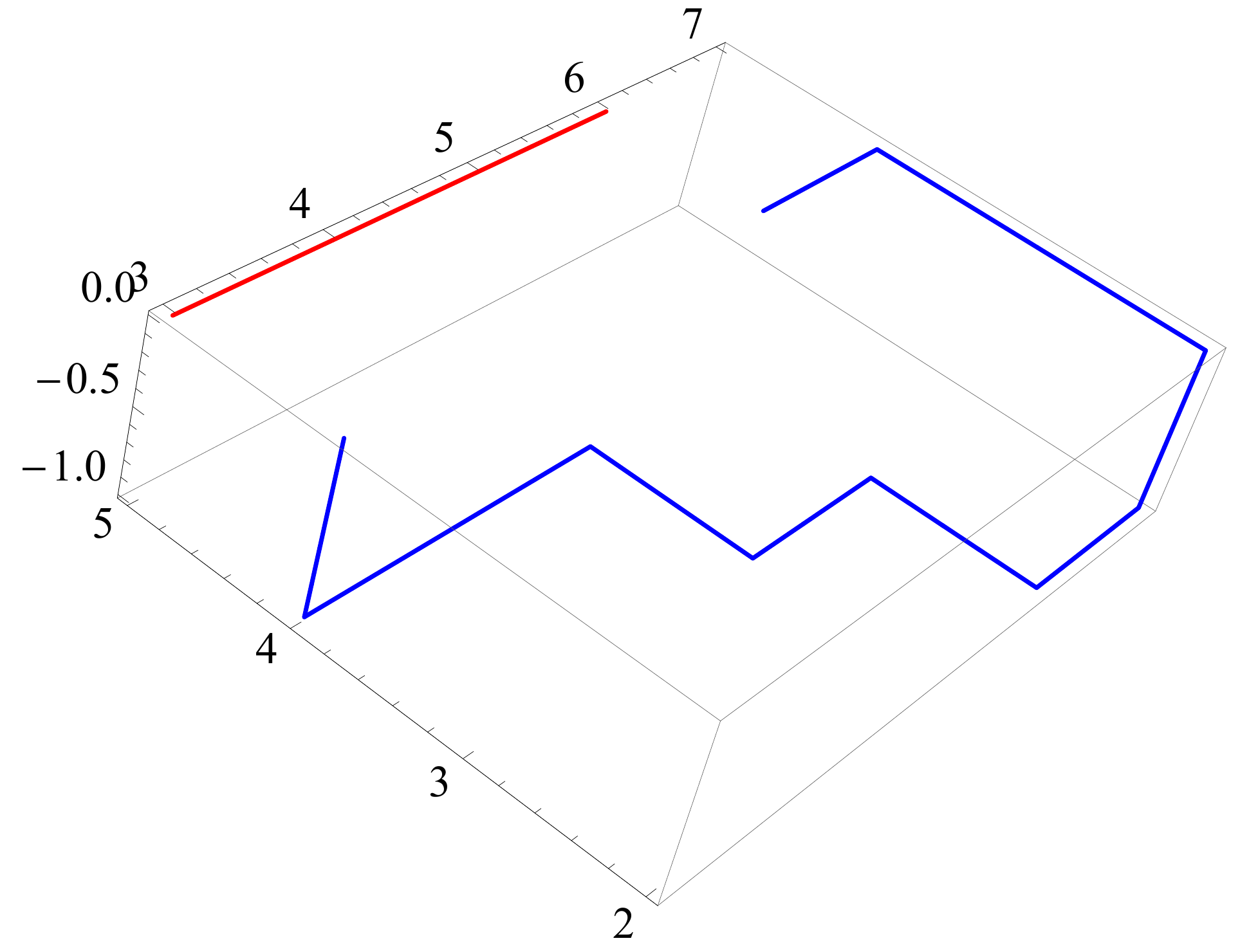}
  \caption{Upper plot: distribution over the string lengths (in units of $a$) in our simulations. The dark (blue) histogram is for $T_0=1\, \GeV,s_T=1.5a$, and the light (orange) one is for  $T_0=1\, \GeV,s_T=1.0 a$.
  The lower plot shows a typical configuration in the second ensemble, with only one string excited.}
  \label{fig_d34_hist}
  \end{center}
\end{figure}

Since in hadronic collisions the color flux conservation requires production of an even number of strings, (most) of our simulations are initialized by the two-string configurations. The end points are separated by a fixed distance $3a\sim 1.5\, \fm$ and are not moved by the update algorithm.

In Fig.~\ref{fig_L(t)_g0} we show an example of history of such simulations, as the string length vs the computer time
$t/t_m$. The time is in units $t_m=10$ of the entire string update cycles. The total run (equilibration time excluded)
is typically about $(1-3) \times 10^4$ iterations.
The necessary run length actually was found to be dependent on the
ball size: the example we will now use corresponds to a ``medium-size ball" with a length of about 50 links and a mass of about 25 $\GeV$.

The integral distribution over all three coordinates is close to the Gaussian one, as is exemplified in the upper figure. Yet it is not
just a Gaussian ensemble of random points, as the points constitute extended objects - strings.  One can see in the lower part of Fig.~\ref{fig_L(t)_g0}
that  the
 (computer time) history of the system displays rather large fluctuations.
 Yet the average over points (not shown) does not show any obvious time dependence, which means that the average properties of the ensemble have stabilized.
The reason for  those is the near-critical conditions at the ball surface, where the string has effectively a very small effective tension.
 Furthermore, if one looks at the individual configurations -- e.g. those displayed in Fig.~\ref{fig_shape} -- one can see that, in spite of relatively heavy string balls,
 most of the space remains unoccupied.

As the parameter $s_T$ of the ball size is reduced, the mean length (and thus the ball's mass) is strongly diminished as well.
Two examples of the length distribution shown in Fig.~\ref{fig_d34_hist} make this point clear.
While at $T_0=1\, \GeV,\, s_T=1.5a$ (dark blue histogram) one finds a string ball of an average length of about 20 links,
 further reduction to $s_T=1.0a$ (light orange histogram) shows that the most probable is the shortest configuration with 8 points (6 links),
 corresponding to an unexcited initial configuration. Yet even in this case, the population of the excited strings still show a long tale, with population up to 25 links (in this simulation), with a probability rate of about a percent.
 Inspection of those configurations shows that it is dominated by the excitation of one of the strings only; see the lower part of Fig.~\ref{fig_d34_hist}.

 \begin{figure}[t]
  \begin{center}
  \includegraphics[width=6cm]{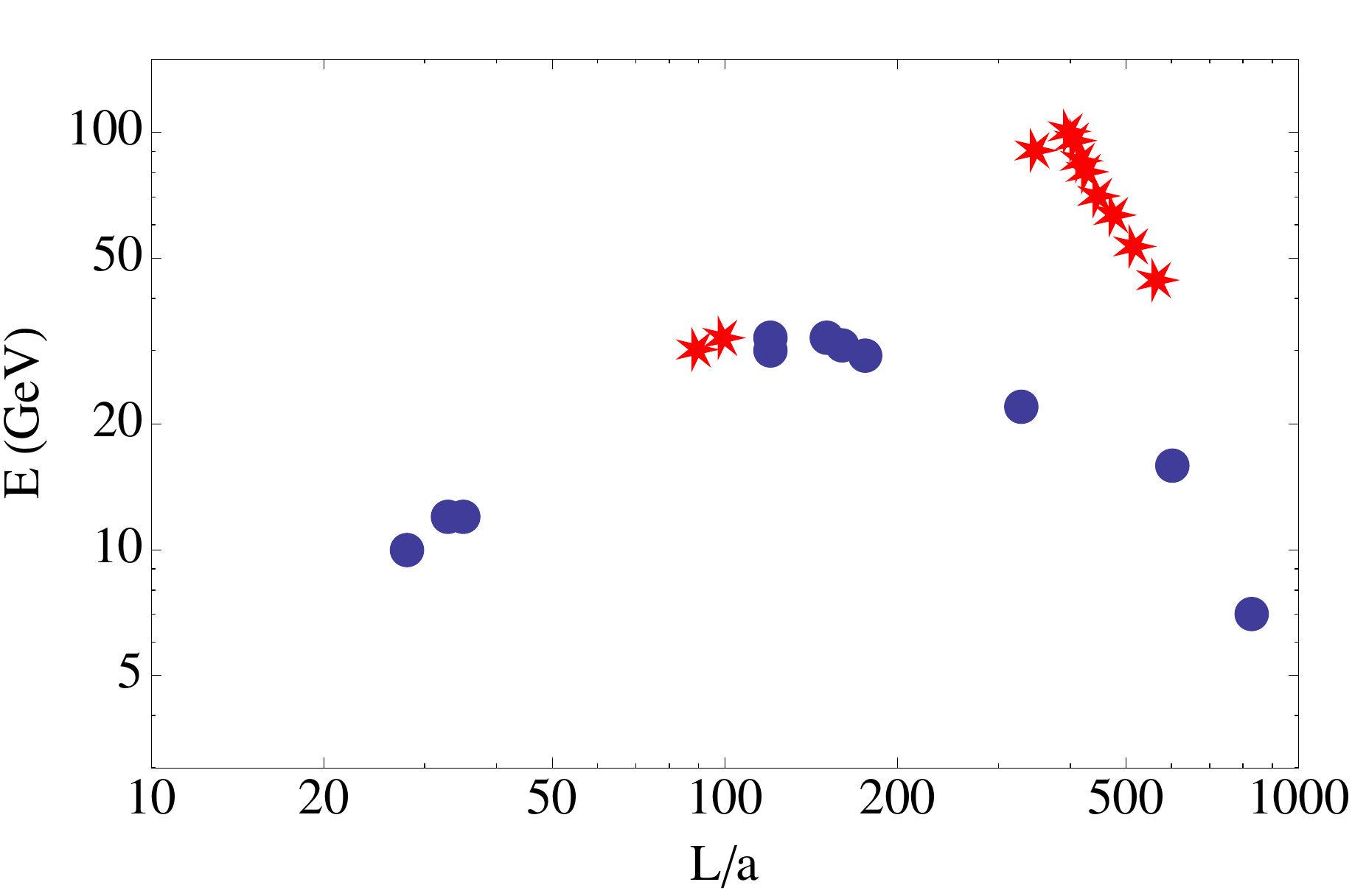}
      \includegraphics[width=6cm]{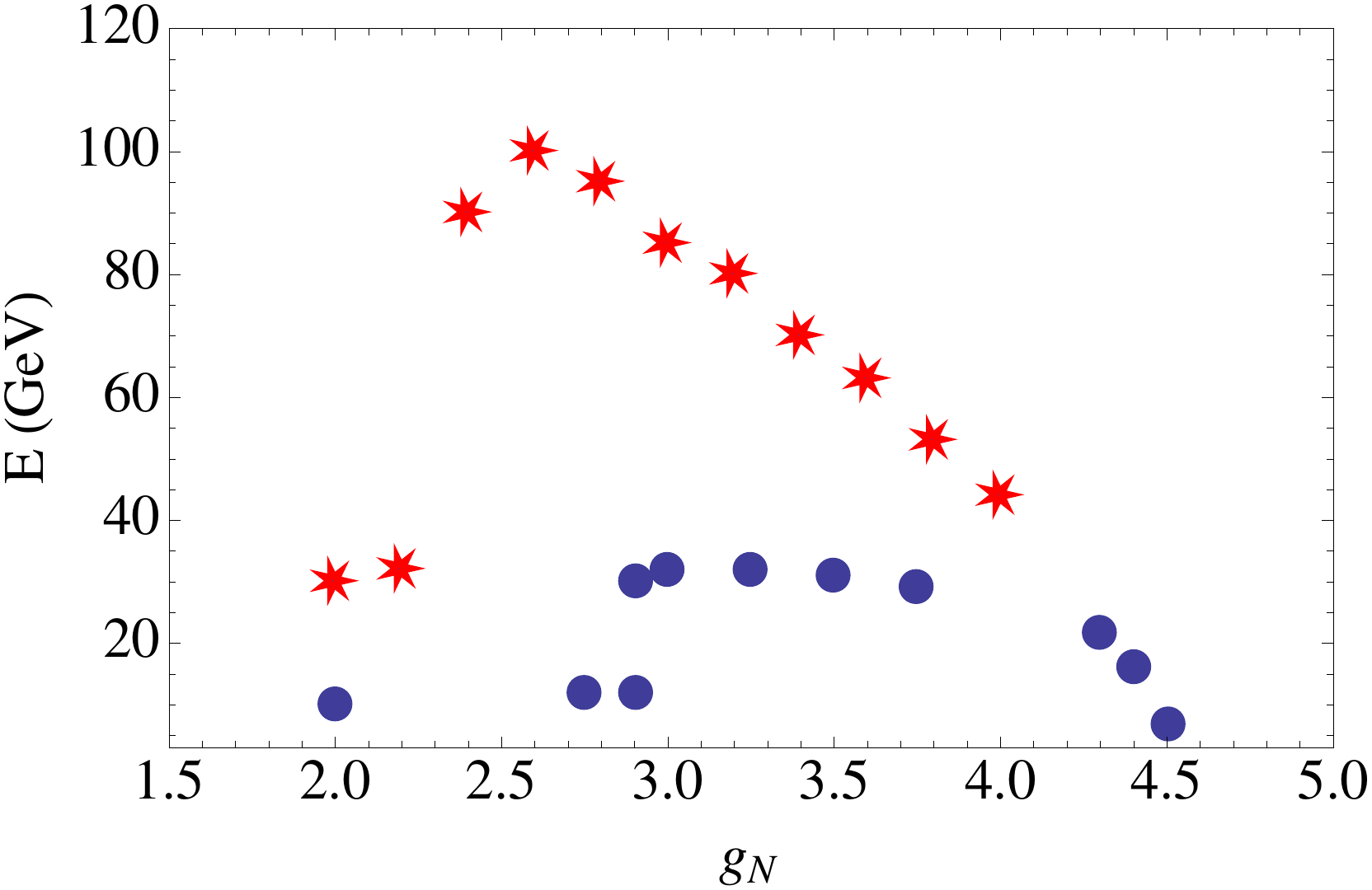}
  \caption{ Upper plot: the mean energy of the cluster $E(g_N)\, [\GeV]$ vs the mean length of the string $L(g_N)/a$.
   Lower plot: the mean energy of the cluster $E(g_N)\, [\GeV]$ vs the ``Newton coupling" $g_N\, [\GeV^{-2}]$.
Points show the results of the simulations in setting  $T_0=1\, \GeV$ and size of the ball $s_T=1.5a, 2a$,
for circles and stars, respectively.
}
  \label{fig_runs0}
  \end{center}
\end{figure}

\subsection{Self-interaction included } \label{sec_with_interaction}

We introduce a Yukawa interaction between string segments, which in our algorithm, for the sake of simplicity, is implemented through an interaction between the string nodes, each of a mass of one string segment, $\sigma_T a$. The Yukawa potential between two string nodes, $\vec r_i$ and $\vec r_j$, is given then by
\begin{align}
V(\vec r_i, \vec r_j) = - \ddd\frac{g_N \left(\sigma_T a\right)^2}{|\vec r_i - \vec r_j|} \exp{\left(-m_\sigma |\vec r_i - \vec r_j|\right)}\,.
\end{align}
This form of the potential is introduced to the update algorithm, i.e., to the probability (\ref{metropolis}).

Now we are ready to see how  nonzero string self-interaction
modifies the properties of the system. While increasing the corresponding parameter --
``scalar Newton's constant''
$g_N$ --
we observe that above its critical value even the most basic features
of the system change.

In Fig.~\ref{fig_runs0} (upper figure), we show the calculated
relation between the average string length $L$ and its energy $E$.
Each point is a run of about 10\textprimstress000 iterations of the entire string updates after equilibration.
 While at small coupling
$E$ and $L$ are simply proportional to each other, like for noninteracting strings described above,
this behavior changes abruptly.  As the negative self-interaction energy
become important, the total energy $E$ of the ball becomes $decreasing$ with the string length  $L$.
 In Fig.~\ref{fig_runs0} (lower figure), we show more details of this behavior: this plot demonstrates how total energy $E$
depends on the coupling value $g_N$.
We find a jump  at the critical coupling (for this setting) $g_N^{c1}$,
which in a simulation looks like a first-order transition, with double-maxima distributions in the energy and length.
As is seen from the figure, the precise value of the coupling somewhat depends on the system size.
At this coupling the jump in energy is always about a factor 3, and the jump in string length (or entropy)
is even larger.

In this way we observe a new regime for our system, which we will call
the ``entropy-rich self-balanced string balls". For a given fixed mass $M$, we thus find
 that string balls may belong to two very distinct
classes: (i) small near-random balls and (ii) large ones in which the string can be very long
but balances its tension by a comparable collective attraction. Discovery of this second regime
 is the main result of this paper.

  \begin{figure}[t]
  \begin{center}
  \includegraphics[width=6cm]{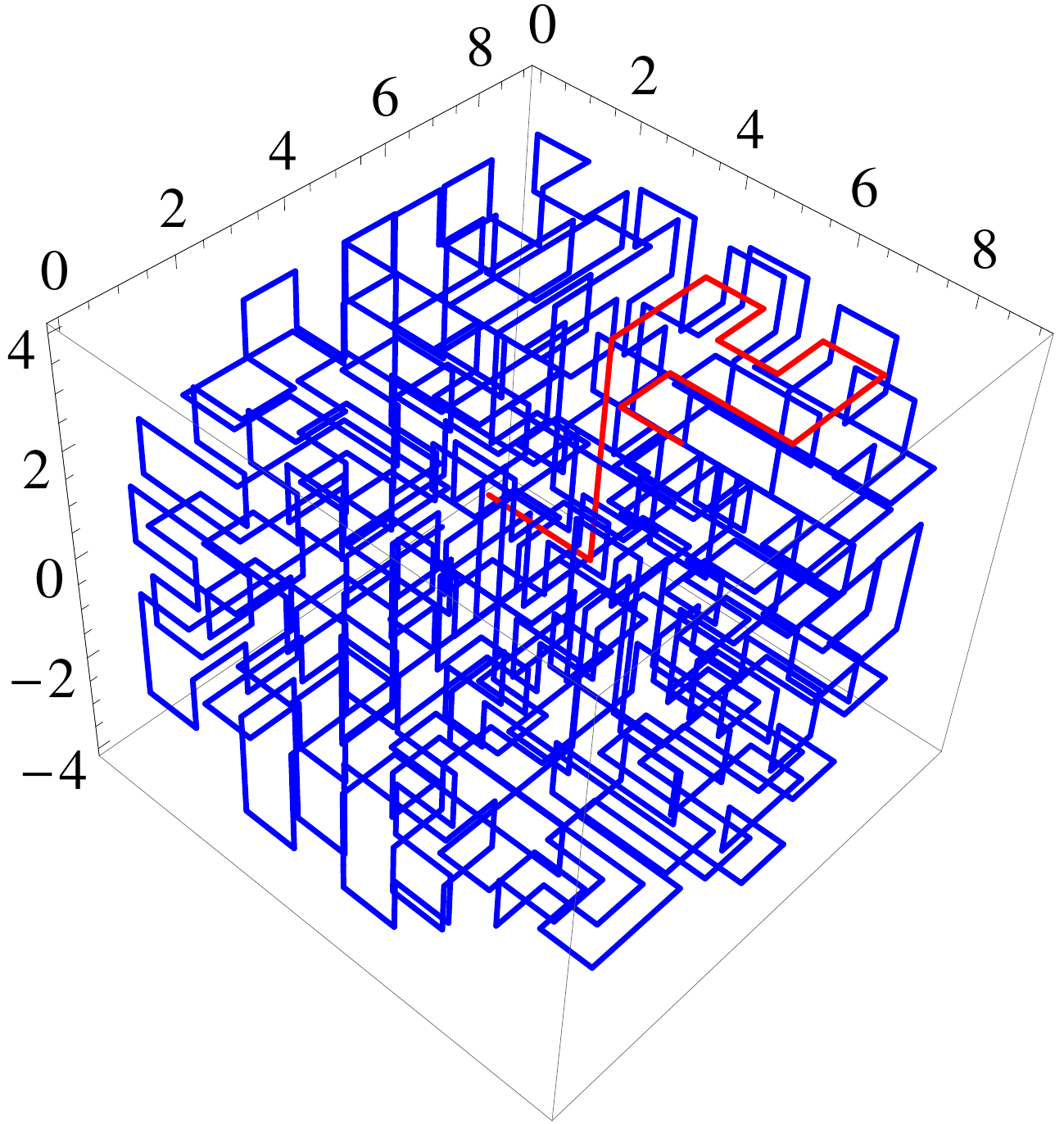}
 \caption{(Color online)
  A typical configuration in the entropy-rich self-balanced string balls ensemble. Simulation
  parameters: $T_0=1\, \GeV,\, s_T=1.5a,\, g_N=4.4\, \GeV^{-2}$. }
  \label{fig_31_config}
  \end{center}
\end{figure}

An example of a corresponding configuration is shown in Fig.~\ref{fig_31_config}. Note that,
in spite of a very large string length $L/a\sim 700$, the total energy is only $E\approx 17\, \GeV$,
as a result of the balancing between the mass and self-interaction.
Note furthermore that that configurations are very asymmetric:
  one string is excited much more than the other, since the longer string has many more states
  than the shorter one. The same feature has been noticed on the lattice as well: typically, one very long string
  forms a large cluster, dominating over a few small clusters.
  Note further that nearly all space inside the ball with $T>T_H$ is occupied.
High entropy corresponds to a (astronomically) large number of shapes
this string may have.

 Finally,
there exists the second critical coupling, which is found to be
$g_N^{c2}\approx 4.5 \, \GeV^{-2}$, above which
balancing the energy becomes impossible and simulations show immediate
collapse of the system, in which the energy quickly falls to large negative values, clearly
of no physical meaning.

\begin{figure}[b]
  \begin{center}
  \includegraphics[width=7cm]{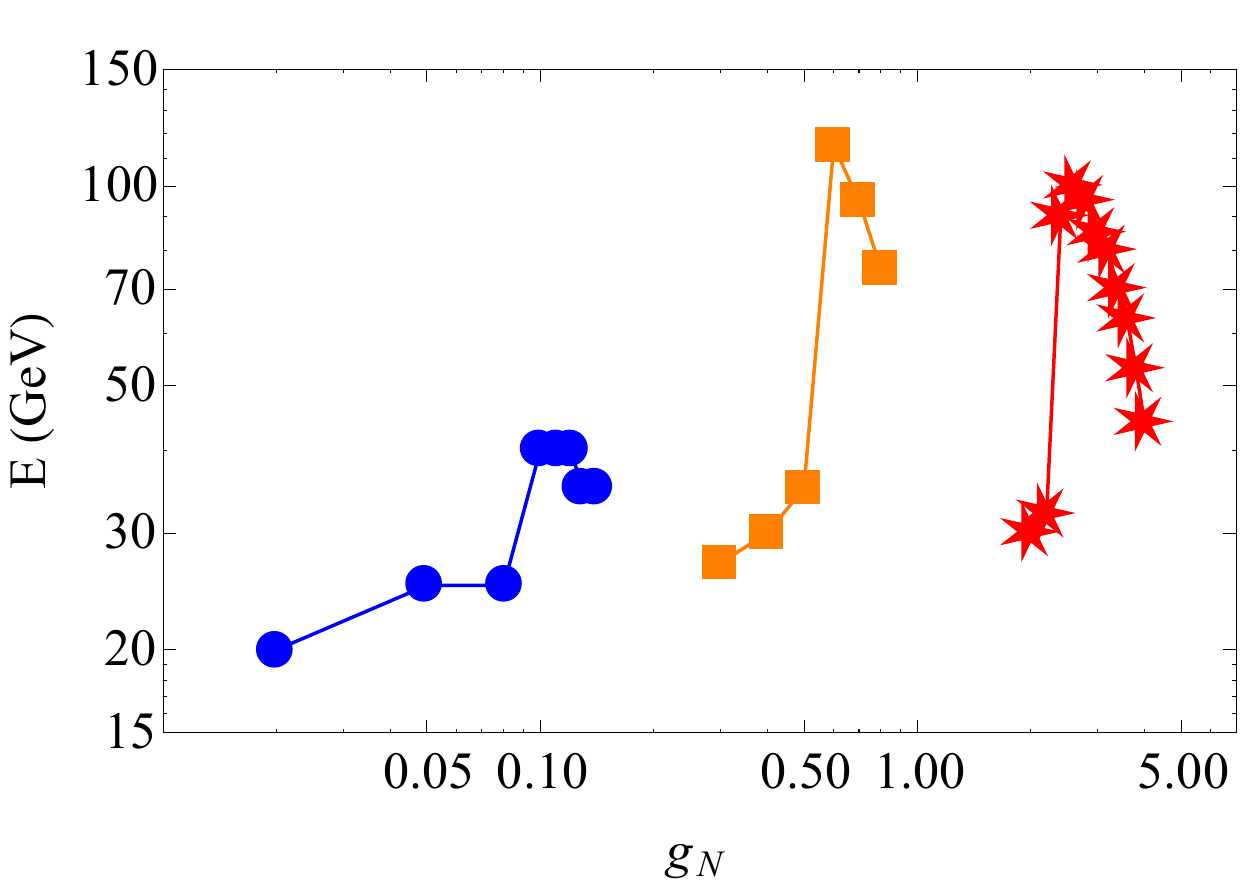}
 \caption{(Color online)
    Blue circles, yellow squares  and red stars show the  dependence of the string ball energy $E (\GeV)$ on the coupling $ g_N \, [\GeV^{-2}]$ for  $m_\sigma= 0.1,0.3, 0.6\, \GeV$, respectively. These simulations are performed in the
 setting  $T_0=1\, \GeV,\, s_T=2a $ . }
  \label{fig_light_sigma}
  \end{center}
\end{figure}

   So far we only used the vacuum value of the sigma meson mass,  $m_\sigma=  0.6\, \GeV$.
   What happens if its value is reduced is shown
in Fig.~\ref{fig_light_sigma}? As one can see from this plot, the critical self-coupling is reduced by about
an order of magnitude between subsequent values of  $m_\sigma$. Indeed, as
the mass decreases by roughly a factor 2,  the volume of the region where $r\leq 1/m_\sigma$ is increased roughly by the factor 8.

As one more remark, in string literature, transition from the strings to black holes is assumed to be smooth.
However, our simulations demonstrated the first-order transition, from the near-free strings to self-bounded string balls
(Fig.~\ref{fig_runs0}). As the mass of the string configurations grows toward the collapse, we expect to approach the black hole parameters. Emission of short strings via the Hawking radiation was not included, since it would complicate the entropy or energy calculation while the strings are very long, and we believe that the total values will not be much affected. Only the total mass
of the ball is relevant.

As a summary of this section, at certain  critical coupling, the string ball undergoes transition to a self-binding
 high-entropy phase. Its value depends strongly on the value of the sigma meson mass at $T_c$.
 (Both the mass and the coupling in the QCD near $T_c$ are not yet known.)

\section{Applications}
\subsection{Jet quenching during the mixed phase   \label{sec_quenching} }
Hard collisions, creating quark and gluon jets, provide an ``x-ray tomography" of the excited fireball
produced in heavy ion collisions. ``Quenching" (absorption, modification) of such jets is one of the
main diagnostic tools used to probe various phases of the hadronic matter appearing
during the fireball expansion. The theory of jet quenching is rather involved, and the phenomenology is even
more complicated, due to the time evolution of the fireball. For a recent summary, see, e.g.,
a report of the JET collaboration \cite{Burke:2013yra} and references therein. For our
purposes it is enough to mention that the relevant matter properties are described by a
single quantity, \be \hat{q}={\dd\langle p_\perp^2 \rangle \over \dd l}\,, \ee
characterizing the increase of the mean squared momentum perpendicular to the direction of motion, per unit length.

   Most early works on the subject assumed that this quantity is simply proportional to the entropy density $s$ of the matter,
   \be {\hat{q} \over s}\approx const\,,   \label{eqn_assumption} \ee
   since both of the have the same mass dimension. Such a naive assumption is reasonable for the QGP
   phase, which is quasiconformal and possesses only one scale -- say $T$-- of its own. But obviously there is no reason
   to extend this assumption to the mixed and hadronic phases, as their structure is quite different,
   especially in respect to the color field distribution affecting $\hat{q}$.
   The characteristic values used in current
   jet quenching models can be seen in Fig.~10 of  Ref.~\cite{Burke:2013yra}: for $T=T_c$ (the mixed phase),
   they range in the following interval:
   \be \left({\hat{q} \over T_c^3}\right)_{min}\approx 1,\qquad  \left({\hat{q} \over T_c^3}\right)_{max}\approx  6\,.\label{JET_minmax} \ee
     Note that the analysis in Ref.~\cite{Burke:2013yra}
   is so far based only on the quenching strength itself; analysis of the quenching for jet paths with different azimuthal angles
   -- or the so-called $v_2=\langle \cos(2\phi) \rangle$ at large $p_T$ --
   is yet to be performed.

    It was,  however, pointed out long ago \cite{Shuryak:2001me} that large experimental
  values of  $v_2$ are difficult to explain by any simple model of quenching; in particular, they were in a strong contradiction with the simplest assumption (\ref{eqn_assumption}).  One possible solution to this puzzle was suggested few years ago in
 Ref.~\cite{Liao:2008dk}: the $v_2$ data can be reproduced, if  $\hat{q}$  is significantly enhanced
  in the mixed phase. More recent data, especially from the LHC, have shown that $v_2$ has, in fact, a rather strong
  $p_T$ dependence and is  decreasing with $p_T$
  of the observed hadron: so the issue seems to exist only for $p_T < 40\, \GeV$ or so.
  Comparison of those data  with various models and discussion can be found
  in  Refs.~\cite{Zhang:2012ie,Betz:2014ooa}.

     Here we want to point out that a natural explanation for the enhanced $\hat{q}$ in the mixed phase can be provided
     by the strings.
   As far as we know, the ``kicks'' induced by the color electric field inside the QCD strings have been ignored in all jet quenching phenomenology; only the fields of ``charges" (quarks and gluons in QGP, hadrons alternatively) were included,
   in the spherical Debye approximation.
   However, if the entire flux of the color-electric field is inside the QCD strings, there are no Coulomb fields of the charges
   and their Debye cloud.
   There are, in fact, two different reasons for it to be the case: (i)  a generic string  enhancement due to the Hagedorn
     phenomenon and (ii) further enhancement  energy due to the string self-interaction, the main subject of this paper.
     We will discuss below those two effects subsequently.

     As we repeatedly emphasized already, in the mixed phase, the strings are close to their
      Hagedorn temperature, so they get easily excited. Let us refer to
     their average length as $\bar{L}$ and to the string radius as $r_s$. The geometrical cross section of the jet-string interaction scales as
   their product; we will use  $2\bar{L} r_s$.

     More accurately, approximating the QCD crossover transition by a first-order transition, one defines
     the mixed phase as $T=T_c\approx 0.17 \, \GeV$ and variable energy (and/or entropy)  density.
      The normalized energy density according to lattice calculations (now for the QCD with quarks, not just for gluodynamics,
      as in Sec.~\ref{sec_lattice}),
   ${\epsilon / T^4}$, ranges from 3 at $T=T_c$ to about 12 at  $T=1.2 \,T_c$.
    Assuming that all this
    energy comes from a string, and dividing naively by the vacuum (T=0) string tension $\sigma_T$, one finds that
   inside each 1 $\fm^3$ cube  there is a string of  length changing between $\bar{L}_{min}=0.4$ and $\bar{L}_{max}= 1.4 \, \fm$, across the mixed phase.

   Let us now estimate $\hat{q}$, by a simple classical argument.
    The  mean square of the momentum kick we write as
   \be \langle p_\perp^2 \rangle \approx(g E r_s)^2\,,  \ee
   which is a color force, $gE$, times the time it acts while the jet is traversing the string, (some coefficient of the order 1 times) $r_s$.
   This combination of the field strength and  the radius can be directly obtained from the following consideration:
   the string tension, i.e., the energy per length, is that of the field inside the string plus the energy of the ``coil" (the magnetic current
   holding the field). The former one is $(E^2/2)\pi r_s^2$, and the latter should be comparable. Assuming it is the same, and eliminating $1/2$,
   we get $\sigma_T= \pi r_s^2 E^2$ from which it follows that
   \be \langle p_\perp^2  \rangle \approx 4\alpha_s \sigma_T \ee
  The geometric probability for a jet to cross the string is
  $ {2 \over 3} {2 \bar{L} r_s \over \fm^2}$ over each $\fm$ longitudinally. Here
  (2/3) excludes string segments along the jet, in which the kick is longitudinal.
   So,
  \be \hat{q} \approx  {16 \over 3} \alpha_s \sigma_T {\bar{L} r_s \over \fm^3}\,.  \label{eqn_profile} \ee
 We still need to know the string radius, and, fortunately,
 its value and the string profile have been extensively studied on the lattice.  Furthermore, in the so-called dual  Abelian model,
 the QCD strings -- flux tubes -- are the well known  Abrikosov
vortex solutions. Numerical data and the dual theory do, in fact, agree quite well; see,
 in particular, a review by Bali \cite{Bali:1998de}, from which we borrow a fit to the lattice data, by the profile function
\be E(x) = {\Phi_e \over 2\pi r_s^2} K_0(x/r_s) \ee
  with $K_0$ being the Bessel function. The main point here is the   value
  of the string radius $r_s=1/(1.3 \GeV)=0.15\, \fm$. The normalization  parameter is $\Phi_e=1.44$.

Now all parameters in the $\hat{q}$ expression above are fixed and we can evaluate $\hat{q}$ numerically.
With $\alpha_s=1/2$ one finds the range across the mixed phase to be
\be \hat{q}_{min}= 0.028  , \,\,\,\quad  \hat{q}_{max}=0.10 \, \left({\GeV^2 \over \fm}\right)\,. \ee
Let us compare these estimates with the values used in the phenomenological models by the JET collaboration (\ref{JET_minmax}) mentioned in the beginning of this subsection. Putting them in the same absolute units, one finds those to be
\be \hat{q}_{min}= 0.025  , \,\,\,\quad  \hat{q}_{max}=0.15 \, \left({\GeV^2 \over \fm}\right)\,, \ee
which is in a good correspondence with our estimates.

This agreement does not, of course, mean that either the estimate or empirical inputs used are, in fact, correct.
Recall  that the JET collaboration's analysis is done for the hadron $p_T\sim 10 \, \, \GeV$, well inside the
region in which the large $v_2$  puzzle remains unresolved. If these data are to be included in their analysis,
the values would go up.

   From the theory side, the presented estimate looks suspicious, because it does not include the second enhancement effect, which is due to the string self-interaction. Indeed, above we
 assumed the energy of the string to be just linear in length due to its (vacuum) tension, i.e., $L\sigma_T$. But, as we demonstrated in the upper Fig.~\ref{fig_runs0}, the
 ``entropy-rich branch" of the string balls has a different relation between the total energy and  the string length $L$:
 self-interaction can compensate a large fraction of the energy.
    For the same total ball energy, its string length $L$ can, in fact, be up to an order of magnitude larger, reaching, perhaps,
    $\hat{q}\sim 1 \, \GeV^2/\fm$ magnitude range, which is usually associated with the QGP phase.
    Since the string inside still contains the same electric flux, etc., it means that   $ \hat{q}$ can be
enhanced by this mechanism by about an {\em order of magnitude}.
            (Another glance at our extreme string-ball configurations shown in  Fig.~\ref{fig_31_config}
            may be needed at this point, for the most skeptical readers.)
If this is the case, the mechanism behind  large $v_2$  will be  explained.

Of course, we treat it just as a possible mechanism, which takes place, if the self-interaction parameter happens to be of the right magnitude.
Unfortunately, we do not really know what its real-world value is.  (Again, we  only see that
what is needed is several times smaller than sigma coupling to the nucleons, which binds them into nuclei.)

\subsection{Angular correlations} \label{sec_angular}
The first difference between the typical and high-multiplicity $pp$ and $pA$ collisions first discovered
was the so-called ``ridge" correlations. Soon LHC experiments had shown that it is, in fact,
complemented by an ``antiridge" out of plane, with
 the second $n=2$ harmonics
  \be v_n=\langle \cos(n\phi)\rangle \ee
in the azimuthal angle.  Also (smaller) third harmonics were observed; both fit well to the hydrodynamic systematics
from $AA$ collisions.

Hydrodynamics explains these azimuthal moments by relatively small deformations
of   basically axially symmetric ($n=0$) flow, known as the  radial flow.
It has been predicted by Shuryak and Zahed \cite{Shuryak:2013ke} that the magnitude of such radial flow
 in $pp$ should be larger than in $pA$, which is larger than ever observed in $AA$ collisions. Using spectra of identified secondaries,
 it has indeed been observed, by  CMS and ALICE collaborations \cite{Chatrchyan:2013eya,Abelev:2013haa}; see also Refs.~\cite{indians}.
 So, supporting
 the hydrodynamical ``explosive" interpretation of high multiplicity $pp/pA$
 events is now becoming a mainstream activity.

  The analysis just mentioned deals with the average values of angular correlation parameters $v_n$; one may also address the question of a $distribution$
 over them. Not going into detailed discussions, we find it
  sufficient to view
hydrodynamics simply by classical means,  translating the initial spatial deformation of the system
\be \epsilon_n = {  \ddd\int \dd^2 r_\perp \cos(n\phi) r_\perp^n (\dd N/\dd^2 r_\perp) \over \ddd \int \dd^2 r_\perp  r_\perp^n (\dd N/\dd^2 r_\perp)} \ee
in the azimuthal angle $\phi$  into the observable flow parameters $v_n$, via certain calculable linear response coefficients,
\be {v_n \over  \epsilon_n}= f_n^{hydro}\,. \label{eqn_lin.responce} \ee
Studies of the $AA$ and $pA$ collisions in the Glauber model provide not only the average means but also
distributions $P(\epsilon_n)$. Observed distributions over $v_n$ are well described by those,
so a linear response relation between them is believed to be valid on the event-by-event basis.

In ``event generators", describing the initial state, one determines $\epsilon_n$ for each event in the ensemble,
obtaining certain distribution of their values. The average over those will be denoted by the angular brackets $\langle ... \rangle$.
The
 experimental measurements involving different numbers of particles are traditionally encoded in the following combinations
 of moments of the distribution (see the derivation and discussion in, e.g., Ref.~\cite{Bzdak:2013rya}),
\begin{align}
\left(\epsilon_n\{2\}\right)^2 =& \langle \epsilon_n^2\rangle\,,  \label{eqn_2468}\\
\left(\epsilon_n\{4\} \right)^4=& 2\langle \epsilon_n^2\rangle^2 - \langle \epsilon_n^4\rangle\,,  \label{eps2-4}\\
  \left(\epsilon_n\{6\} \right)^6=& {1 \over 4}\left[\langle \epsilon_n^6\rangle -9\langle \epsilon_n^2\rangle\langle \epsilon_n^4\rangle + 12\langle \epsilon_n^2\rangle^3\right]\,,\\
\left(\epsilon_n\{8\} \right)^8=&{1 \over 33}\left[ -\langle \epsilon_n^8\rangle + 16\langle \epsilon_n^6\rangle\langle \epsilon_n^2\rangle + \right. \\
 &\left. 18\langle \epsilon_n^4\rangle^2 -144\langle \epsilon_n^4\rangle\langle \epsilon_n^2\rangle^2 + 144\langle \epsilon_n^2\rangle^4\right]\,.
\end{align}
Assuming the linear response (\ref{eqn_lin.responce}) holds on the event-by-event basis, one finds that it will also hold
for $v_n$ distributions, as all terms in the formulas above scale in the same way.

 The corresponding combinations are called similarly $v_n\{m\}$; the parameter $m$ in this case has a meaning of the
 number of secondaries used to produce this correlation. We  will make two brief comments on those parameters.
 One is that ``classical" (=nonfluctuating) distribution $P(v_n)=\delta(v_n-V_n^0)$ makes all of them be the same, namely, $V_n^0$.
However, the
empirical pattern observed in $AA$ and $pA$ collisions is a bit different, i.e.
\be v_2\{2\}> v_2\{4\}\approx v_2\{6\}\approx v_2\{8\}\,,    \label{eqn_pattern} \ee
which implies that there exist certain two-body correlations absent for higher $m>2$. We will not go into the further
 discussion present in the literature of what exactly it tells us about the underlying physics.

While fluctuations in the location of the nucleons are believed to be the main cause of angular shape
fluctuations -- thus, Glauber treatment -- of $AA$ and $pA$ collisions, so far there is no consensus or even studies
of what can be the source of those in $pp$ collisions. So far, studies of the proton structure
in terms of parton distribution functions were restricted to their longitudinal momenta only,
and also to minimal biased (mean) protons. Only qualitative speculations on the transverse
shapes of the high multiplicity events were proposed: for example, one suggestion made by Bjorken \textit{et al.} \cite{Bjorken:2013boa} is that
such events come from the collisions of near-parallel strings.
If so, one may expect very large initial deformations $\epsilon_2\approx 1$, but $\epsilon_3\approx 0$.

Since ensembles of string balls generated by our partition function have various angular shapes,
and since some of those may be driven by the underlying explosive hydro (as explained in the introduction),
some study of their angular fluctuations and statistical ensembles involved are of a certain interest.
Note that, because points on our lattices are connected by the string, their statistics may be
different from the ensembles of mutually  independent points (the ``wounded nucleons")
in the Glauber approach.

\begin{figure}[t]
  \begin{center}
  \includegraphics[width=6cm]{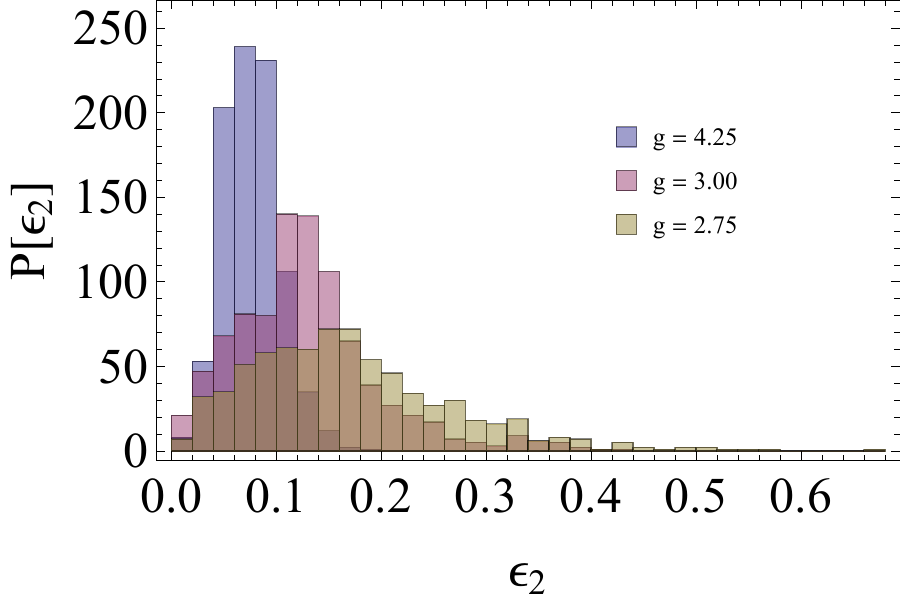}
      \includegraphics[width=6cm]{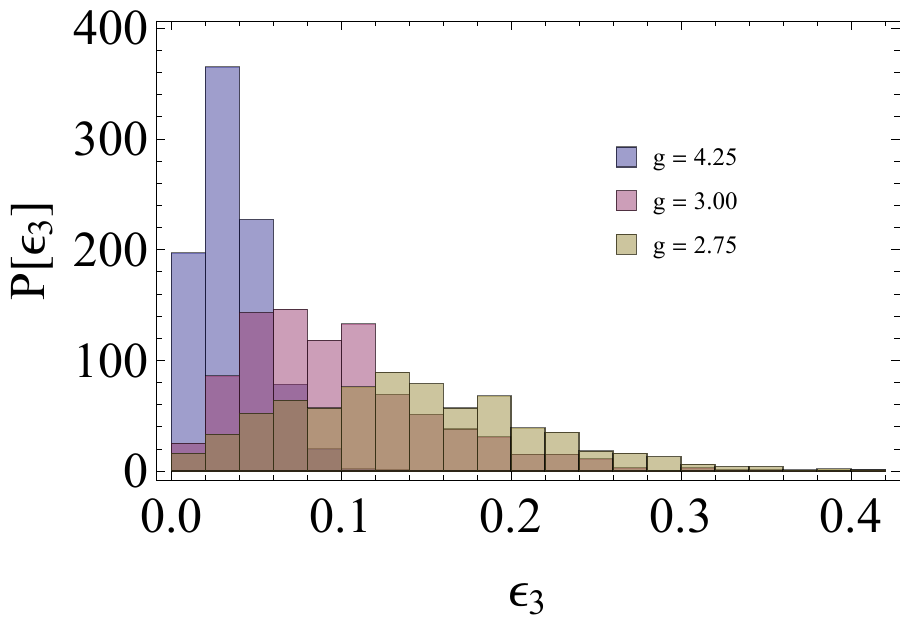}
  \caption{ The distributions over $\epsilon_2$ and $\epsilon_3$ (upper and lower plots), for several
  values of the  ``Newton coupling" $g_N\, [\GeV^{-2}]$.
}
  \label{fig_v2v3}
  \end{center}
\end{figure}

In Fig.~\ref{fig_v2v3} we show the (non-normalized) distributions $P(\epsilon_2),P(\epsilon_3)$ for three ensembles of string balls.
The first observation following from those distributions is that the larger are the balls, the smaller are their deformations.
This is, of course, expected on the general grounds; note, however, that for our largest ball the peak values are at
$\epsilon_2\approx 0.06, \epsilon_3\approx 0.03$, which are quite small, in the ballpark of those in $AA$ collisions with heavy nuclei.
  The second qualitative feature of the distributions is that they are quite far from the delta function or narrow Gaussian: in fact, there exists a
  rather long tail toward large values of deformations. This is a consequence of near-critical fluctuations of strings
  on the ball surface, allowing for a rather large string suddenly protruded outward (as, e.g., is the case for
  magnetic flux tubes on the Sun).
  The third feature is more developed tails for $\epsilon_3$ than for $\epsilon_2$, in difference to Glauber ensembles,
  which show rather similar fluctuations in  all $\epsilon_n$ up to high $n$, because of the pointlike  angular
  contributions of the ``wounded nucleons". Our strings are not pointlike objects, and this difference, indeed, reveals itself.

  The specific values of the parameters (for the $g_N=4.25\, \GeV^{-2}$, the largest used) are
  \begin{align}
  \epsilon_2\{2\}=0.0759,\qquad & \epsilon_2\{4\}=0.0621,\\
  \epsilon_2\{6\}=0.0636,\qquad & \epsilon_2\{8\}= 0.0635,
 \end{align}
 which indeed fits the pattern (\ref{eqn_pattern}), whatever it may mean. However, some of the smaller balls
 have ensembles which fluctuate so much as they violate the indicated pattern, sometimes to the extent that  the rhs of
  Ref.~(\ref{eps2-4}) becomes negative, or $\epsilon_2\{4\}$ complex.

  We are looking forward toward
  experimental measurements of those distributions/parameters in the high multiplicity $pp$ collisions.

\section{Summary and Outlook}
\subsection{Summary} \label{sec_summary}
  In this paper we have formulated a new lattice string model, reminiscent of Wilson's strong coupling
  expansion idea. Its lattice spacing $a$ is not, however, an artificial discretization parameter, going to zero
  at the end of the calculation. Instead, it is chosen to imitate a physical density of states of the QCD string.

  Our setting  also differs from the lattice gauge theories in the infrared. Instead of a certain periodic box, we
  have chosen to have an infinite lattice, but with a space-dependent temperature $T(x)$.
  The resulting string balls of a certain predetermined size are expected to model excited systems created both
  at the mixed phase of the heavy ion collisions and also in some high multiplicity $pp$ scattering events.

  The main physics of the model is related to the string self-interaction. While this has been discussed
  in the fundamental string theory context --  with the self-interaction due to gravity, leading sometimes to the creation of black
  holes -- we believed it has never been discussed for QCD strings. In particular, various event generators
  widely used by experiments all follow the Lund model treatment of strings; i.e., they are
  simply straight in shape and noninteracting.

  The main result of the simulation we made in the model's setting is a
  discovery of a new class of string balls, in which energy of the string is balanced by the
  self-interaction. Such objects may form a relatively modest mass with rather large entropy,
   and thus provide the closest QCD-based objects to the gravitational black holes.

   Obviously, those objects are of significant theoretical interest. After all, fundamental string theory
  is still remote from experiments, by many orders of magnitude, and small-mass gravitational
  black holes are unavailable. QCD systems we discuss are, on the other hand, produced in the numbers of billions
  at hadronic and heavy ion colliders, providing statistically interesting samples of many categories
  of the final states. Strong collective fields, if they exist in some of those systems, must in particular
  give rise to the massive pair production at their edge, analogous in spirit to the Hawking radiation.
  Large entropy and its possible upper bounds should be in a way analogous to the Bekenstein
  entropy of the black holes. The impossibility of climbing out of a deep potential well, gravitational or not,
  causes the famous information paradox, which has been hotly debated in the literature for a long time.

  Furthermore, as we comment on a bit in the next subsection, various versions of AdS/QCD
   models bring QCD string balls and gravitational black holes together in a rather direct sense via the
   holography.

   String systems with large entropy are also of significant experimental interest. As entropy  never decreases,
   those must be related to a larger entropy at later hadronization stages and thus can be detected
   via unusually large multiplicities. As we commented on in the introduction, LHC experimental triggers were
   able to select  statistically significant ensembles of such events in $pp$, $pA$ collisions and demonstrate that their behavior
   is quite different from the minimal biased or ``usual'' events.
    In the previous section, we, in particular, argued that the high multiplicity events correspond to the high-entropy string balls;
    one consequence of this is a relatively small angular deformation $\epsilon_2$.

\subsection{Outlook} \label{sec_outlook}

One application of the self-interacting strings has been developed in our other paper \cite{Kalaydzhyan:2014zqa}
in which we discussed a ``spaghetti" configuration, made of parallel strings. Such
configuration arises due to the strings being stretched along the beam direction in high-energy collisions.
We found at which number of strings one finds a ``spaghetti collapse" and argue that this is
indeed what happens in central $pA$ and peripheral $AA$ collisions.

 At LHC the fraction of explosive events in $pp$ and $pA$ collisions has a very small
 probability $\sim 10^{-6}$ \cite{Khachatryan:2010nk, Chatrchyan:2013nka}. In Ref.~\cite{superhigh} we further noted that at ultra-high energy
 collisions, corresponding to the highest-energy cosmic rays observed, e.g. at Pierre Auger observatory, the rate of such
 events should become large $\mathcal{O}(1)$, affecting, in particular, the mean $\langle p_T \rangle$ and the angular distribution of produced particles.

As we already emphasized in the introduction, one rather straightforward extension of this work can be performed within the holographic AdS/QCD framework.
This implies that one of the three spatial dimensions of the model is interpreted as a
holographic curved coordinate $z$. The bulk fields, dilaton and graviton, are massless,
but their motion in $z$ gets quantized due to a confining wall, and the corresponding
5-momenta play the role of the sigma mass in our Yukawa potential.

   Perhaps more demanding in terms of work would be its extension from the early time -- when string
   configurations are quantum fluctuations in transverse coordinates described by the semiclassical ``effective temperature" -- to later times.  An individual QCD string, from being transverse at the initial collision time $t=0$, gets stretched longitudinally and then breaks, first into clusters and then to final  lightest hadrons reaching the detectors.
    Stretching of a self-interacting string ball is an interesting problem, not yet studied.

    The holographic strings cannot be broken -- as considered AdS/QCD models lack fundamental charges in the bulk --
    but simply fall to the ``bottom of space", i.e. off the boundary,
    influencing holographically the stress tensor and other operators on the boundary.
          A holographic string ball, if heavy enough,
    makes a full-scale bulk black hole, with all its prerequisites -- Hawking-Bekenstein temperature and entropy,
    as well as gravitational fall away from the boundary, dual to hydro explosion.
    A version of it, also for the spaghetti configuration, is considered in Ref.~\cite{adsqcd}.

\vskip 1cm
{\bf Acknowledgements.}
The main acknowledgement goes to Ismail Zahed, who worked on versions of holographic Pomeron
and QCD strings for many years; this work would never have been done without his  input.
Section \ref{sec_quenching}  originated from a suggestion by Dima Kharzeev, in a discussion with Jinfeng Liao.
We would also like to thank  Gokce Basar  and Derek Teaney for multiple useful discussions and
Bill Zajc, who  read version 2 of the paper and provided many useful comments.
This work was supported in part by the U.S. Department of Energy under Contract No. DE-FG-88ER40388.
The calculations were performed partially at the DESY theory clusters.


\end{document}